\documentclass[twocolumn]{aastex61}

\usepackage{graphicx,amsmath,amsfonts,amssymb,subfigure,natbib,hyperref}
\newcommand{\mname}{\texttt{Prospector-$\alpha$}}

\newcommand{\prospector}{\texttt{Prospector}}

\newcommand{\fagn}{f$_{\mathrm{AGN}}$}
\newcommand{\fmir}{f$_{\mathrm{AGN,MIR}}$}
\newcommand{\tauagn}{$\tau_{\mathrm{AGN}}$}
\newcommand{\halpha}{H$\alpha$}
\newcommand{\hbeta}{H$\beta$}

\newcommand{\dn}{D$_{\mathrm{n}}$4000}

\newcommand{\umin}{U$_\mathrm{min}$}
\newcommand{\gammae}{$\gamma_{\mathrm{e}}$}
\newcommand{\qpah}{Q$_{\mathrm{PAH}}$}
\newcommand{\herschel}{\textit{Herschel}}
\newcommand{\spitzer}{\textit{Spitzer}}
\newcommand{\angstrom}{\mbox{\normalfont\AA}}
\newcommand{\wise}{\textit{WISE}}
\newcommand{\wisegradient}{$\nabla$(W1-W2) at 2 kpc}

\newcommand{\lir}{{L$_{\mathrm{IR}}$}}
\newcommand{\luv}{{L$_{\mathrm{UV}}$}}

\begin{document}

\title{Hot dust in Panchromatic SED Fitting: Identification of AGN and improved galaxy properties}

\correspondingauthor{Joel Leja}
\email{joel.leja@cfa.harvard.edu}

\author[0000-0001-6755-1315]{Joel Leja}
\affil{Harvard-Smithsonian Center for Astrophysics, 60 Garden St. Cambridge, MA 02138, USA}
\affil{NSF Astronomy and Astrophysics Postdoctoral Fellow}

\author[0000-0002-9280-7594]{Benjamin D. Johnson}
\affil{Harvard-Smithsonian Center for Astrophysics, 60 Garden St. Cambridge, MA 02138, USA}

\author[0000-0002-1590-8551]{Charlie Conroy}
\affil{Harvard-Smithsonian Center for Astrophysics, 60 Garden St. Cambridge, MA 02138, USA}

\author[0000-0002-8282-9888]{Pieter van Dokkum}
\affil{Department of Astronomy, Yale University, New Haven, CT 06511, USA}

%\author{Joel Leja\altaffiliation{1}, \altaffiliation{1}, Charlie Conroy\altaffiliation{1}, Pieter G. van Dokkum\altaffiliation{2}}

%\altaffiltext{1}{}
%\altaffiltext{2}{}
%\altaffiltext{3}{Department of Astronomy, University of Washington, Seattle, WA 98185, USA}

%%%%%%%%
% ABSTRACT %
%%%%%%%%

\begin{abstract}
Forward modeling of the full galaxy SED is a powerful technique, providing self-consistent constraints on stellar ages, dust properties, and metallicities. However, the accuracy of these results is contingent on the accuracy of the model. One significant source of uncertainty is the contribution of obscured AGN, as they are relatively common and can produce substantial mid-IR (MIR) emission. Here we include emission from dusty AGN torii in the \prospector{} SED-fitting framework, and fit the UV-IR broadband photometry of 129 nearby galaxies. We find that 10\% of the fitted galaxies host an AGN contributing $>$10\% of the observed galaxy MIR luminosity. We demonstrate the necessity of this AGN component in the following ways. First, we compare observed spectral features to spectral features predicted from our model fit to the photometry. We find that the AGN component greatly improves predictions for observed \halpha{} and \hbeta{} luminosities, as well as mid-infrared {\it Akari} and {\it Spitzer}/IRS spectra. Second, we show that inclusion of the AGN component changes stellar ages and SFRs by up to a factor of 10, and dust attenuations by up to a factor of 2.5. Finally, we show that the strength of our model AGN component correlates with independent AGN indicators, suggesting that these galaxies truly host AGN. Notably, only 46\% of the SED-detected AGN would be detected with a simple MIR color selection. Based on these results, we conclude that SED models which fit MIR data without AGN components are vulnerable to substantial bias in their derived parameters.
\end{abstract}

\section{Introduction}
Fitting physical models to the spectral energy distributions (SEDs) of galaxies is the foundation of modern galaxy evolution studies (see \citealt{walcher11,conroy13a} and references therein). Early versions of galaxy SED models were mostly limited to fitting stellar emission in the broadband UV-NIR \citep{bolzonella00,brinchmann00,papovich01,salim07,brammer08,kriek09}, though there are a few notable exceptions \citep{silva98,devriendt99}. Such models are able to constrain stellar masses to within a relative error of $0.1-0.2$ dex \citep{bell01,wuyts09,muzzin09}. However, they provide poor constraints for dust attenuation and star formation histories \citep{papovich01,shapley06,kriek08,taylor11,wuyts11a}. As a result, star formation rates (SFRs) measured from broadband photometry fit with these models are systematically biased in galaxies with significant dust attenuation or low levels of star formation \citep{wuyts11a,belli17}.

Self-consistent models for the full UV-IR SED of galaxies have been developed in recent years. Full SED models remove these biases by directly modeling the dust emission of galaxies \citep{burgarella05,dacunha08,groves08,noll09,leja17}. Full SED models assume energy balance, where the stellar energy attenuated by dust is then re-emitted in the IR \citep{dacunha08}. In this way, the combination of observed MIR and FIR photometry and a full SED model provide critical new constraints on the total energy budget of galaxies. Full SED physical models produce SFRs consistent with more expensive spectroscopic emission line measurements \citep{shivaei16a,leja17}, validating the accuracy of the dust attenuation and the amount of dust-obscured star formation derived from these models. Furthermore, a complete picture of the galaxy dust and energy budget provides unique constraints on new galaxy physical parameters, such as the shape of the dust attenuation curve to the stellar metallicities \citep{leja17}.

However, incorporating MIR photometry into SED models also introduces sensitivity to new systematics, including AGB circumstellar dust emission \citep{villaume15} and, more saliently, the MIR emission from AGN. AGN can emit significant fractions of their energy in the MIR by heating nearby dust to high temperatures \citep{nenkova08a,padovani17}. Full SED models will naively attribute AGN emission to dust heated by star formation, resulting in full SED models overestimating sSFRs by up to 0.6 dex for galaxies with AGN \citep{salim16}. This empirical finding is consistent with mock tests of SED fitting procedures, which find that the presence of a buried AGN can bias star formation rate estimates even at low AGN contributions of L$_{\mathrm{AGN}}$/L$_{\mathrm{galaxy}} \approx 0.1-0.2$ \citep{ciesla15b}.

The importance of accounting for AGN emission is underscored by the fact that AGN are known to exist in a large fraction of the galaxy population. 35\% of SDSS galaxies detected in all four requisite emission lines have composite or AGN classifications in the Baldwin-Phillips-Terlevich (BPT) diagram \citep{kauffmann03b}. Studies that incorporate radio, X-ray, and MIR detection methods for AGN suggest that 37\% of star-forming galaxies at $0.3 < z < 1$ host an AGN \citep{juneau13}. The prevalence of AGN may increase with redshift as well: \citet{kirkpatrick15} find that up to 40\% of 24$\mu$m selected galaxies host an AGN. A redshift-dependent density of AGN will produce a redshift-dependent systematic bias in galaxy properties. These redshift-dependent biases introduce serious systematics into attempts to track the mass evolution of the galaxy population through time \citep{leja15} and confuse comparisons of observed galaxy properties with numerical simulations of galaxy formation (e.g., \citealt{genel14,furlong15}).

To address these issues, here we include AGN templates in the full SED fitting framework \mname{} described in \citet{leja17}. We validate the accuracy of the recovered galaxy parameters from this new model by comparing to the detailed spectrophotometric catalog of \citet{brown14}, and also demonstrate that the adopted AGN templates are successful in primarily identifying AGN. Section \ref{sec:data} introduces the photometry and spectra used in this study. Section \ref{sec:sedmodel} describes the  \mname{} model and the new AGN templates. Section \ref{sec:results} discusses how the photometric residuals improve when the AGN templates are adopted, and compares the photometry-based predictions for the observed MIR spectra, \halpha{}, and \hbeta{} luminosities for models with and without AGN. Section \ref{sec:broader_galeffect} shows which galaxies host buried AGN, and how the AGN templates improve the accuracy of the recovered host galaxy properties. Section \ref{sec:agn} investigates whether the AGN template is identifying hot dust heated by AGN or by some other mechanism, by comparing the \mname{}-identified AGN population to external indicators of AGN activity. The discussion and conclusion are found in Section \ref{sec:discussion} and Section \ref{sec:conclusion}. This work uses a \citet{chabrier03} initial mass function (IMF) and a WMAP9 cosmology \citep{hinshaw13}.

\section{Data}
\label{sec:data}
\subsection{Photometry and Spectra}
We fit galaxies from the \citet{brown14} spectrophotometric catalog. This catalog includes 129 galaxies within the local universe, all within 250 Mpc. The sample selection is not volume-complete but instead is based on the availability of aperture-matched optical spectra. It covers a diverse set of galaxies, from star-forming low-mass dwarf galaxies to dusty LIRGS and ULIRGS to massive quiescent galaxies. This is a well-studied sample with no strong evidence for quasar-like power-law emission in the UV.

There are 26 bands of broadband photometry available, covering the far-UV to the mid-infrared. These include images from {\it Swift}/UVOT, {\it Galaxy Evolution Explorer} (GALEX), the Sloan Digital Sky Survey (SDSS), the Two Micron All Sky Survey (2MASS), \spitzer{} IRAC and MIPS 24$\mu$m, and the {\it Wide-field Infrared Space Explorer} (\wise{}). For 26 of these galaxies, \herschel{} PACS and SPIRE imaging from the KINGFISH survey \citep{kennicutt11} is also included.

An error floor of 5\% of the flux in each band is enforced, to allow for potential systematic errors both the photometric zero-points and in the physical models for stellar, gas, and dust emission. Additionally, a 30\% error floor is enforced for the \wise{} W3 photometry, as it is subject to a deep, highly variable 10$\mu$m silicate absorption feature which is not included in our models (see Section \ref{sec:agntemplates} for further discussion).

The \citet{brown14} catalogs also provide optical spectra from multiple ground-based telescopes, mid-infrared \spitzer{} spectra \citep{houck04}, and {\it Akari} spectra \citep{onaka07}. The optical spectra are sourced from \citet{moustakas06}, \citet{moustakas10}, \citet{kennicutt92} and \citet{gavazzi04}, and have a resolution of $R \sim 650$ and a wavelength coverage of 3650 to 6900 \angstrom{}. The optical spectra are aperture-matched to the photometry, so no aperture corrections are necessary.

Key spectral features are measured from the optical spectra as described in \citet{leja17}, including emission line luminosities and \dn{}. The measurement described in \citet{leja17} has been updated to include the uncertainty in the model continuum, by drawing model spectra randomly from the \mname{} posteriors. This is particularly important for galaxies with weak \halpha{} and \hbeta{} emission, where the total emission line flux is highly dependent on the (unknown) depth of the underlying stellar absorption. These measurements are compared to predictions from fits to the photometry. The spectra are {\it not} fit directly.

\subsection{{\it WISE} Images}
MIR color gradients for galaxies from the \citet{brown14} catalog are measured from images taken by \wise{}. The official \wise{} image reductions are intentionally convolved with the point-spread function (PSF), which causes an unnecessary loss of information when measuring the color gradients of resolved galaxies. To avoid this, we use reduced images from the \citet{lang14} {\it unWISE} catalog, which preserve the native resolution of the \wise{} images.

We download the {\it unWISE} W1 3.3$\mu$m and W2 $4.6\mu$m images for each galaxy, and convolve the W1 image to the W2 resolution. The background for both images is measured via iterative sigma-clipping and subtracted, and a W1-W2 color map is produced. The centroid of the W1 intensity map is taken as the center of the galaxy, and the W1-W2 color within a circular aperture of $r=2$ kpc and a circular annulus of $2<r<4$ kpc is measured. The color gradient, $\nabla$(W1-W2) [2kpc], is calculated from these measurements. The \texttt{photutils} \citep{bradley16} and \texttt{astropy} \citep{astropy13} python packages are used for this analysis.

We remove galaxies with $\sigma$(W1-W2) $>$ 0.25 mag arcsec$^{-1}$ and galaxies not resolved to within 2 kpc, given the \wise{} PSF. We also remove galaxies where more than 20\% of the pixels in the larger aperture are background pixels. This leaves measurements for 81 galaxies.

\subsection{X-Ray fluxes}
\label{sec:xraytables}
We retrieve Chandra X-ray fluxes for the galaxies in the \citet{brown14} sample from the Chandra Source Catalog (CSC; \citealt{evans10}), the Chandra ACIS Survey of Nearby Galaxies X-Ray Point Source Catalog (CHNGPSCLIU; \citealt{liu11}), and the Chandra XAssist Source List (CXO; \citealt{ptak03}). To be paired with a galaxy, we require X-ray sources to be within 30$"$ of the optical center of the galaxy. The CSC and CHNGPSCLIU catalogs distinguish between extended emission and point sources; all extended emission matches are removed. If multiple X-ray sources are returned, the brightest source is taken as the match.
\begin{figure*}[t!h!]
\begin{center}
\includegraphics[width=0.95\linewidth]{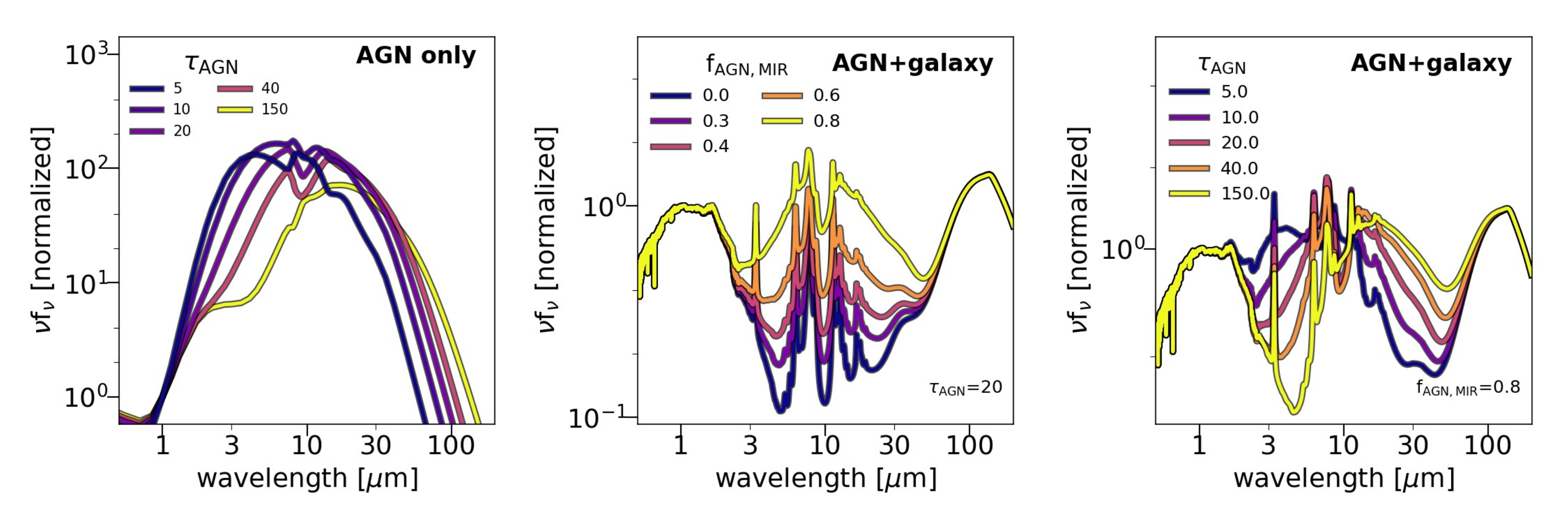}
\caption{AGN models from \citet{nenkova08a} and their implementation in \mname{}. On the left are pure \citet{nenkova08a} AGN templates. They are parameterized by \tauagn{}, the optical depth of an individual dust clump at 5500 \AA{} in the dusty AGN torus. The right two panels contain the model response to the two AGN parameters adopted in \mname{}, on top of an underlying galaxy SED with a constant star formation history over 13.8 Gyr, $A_V=1$, and a typical PAH contribution. The two adopted AGN parameters in \mname{} are normalization \fmir{}, defined as the fraction of observed 4-20$\mu$m flux emitted by the AGN, and \tauagn{}.}
\label{fig:sedresponse}
\end{center}
\end{figure*}

These databases provide fluxes in slightly different Chandra energy bands (0.5-7 keV for CSC, and 0.3-8 keV for CHNGPSCLIU and CXO). We correct to a common 0.5-8 keV window by assuming the standard AGN power-law spectrum:
\begin{equation}
n(E) dE \propto E^{-\Gamma} dE
\end{equation}
where the photon index, $\Gamma$, is taken to be -1.8 \citep{mushotzky93}. These corrections are 1.01 for CSC, and 0.64 for CHNGPSCLIU and CXO.

\section{SED Modeling}
\label{sec:sedmodel}
In this section we briefly describe the \mname{} galaxy SED model, and introduce the FSPS adaptation of the CLUMPY AGN templates from \citet{nenkova08b}.

\subsection{\mname{}}
We adopt the \mname{} SED model as described in \citet{leja17}. This model is implemented in the \prospector{} inference framework (Johnson et al., in preparation), which uses a Bayesian MCMC approach to modeling galaxy SEDs. In brief, the \mname{} model includes a 6-component non-parametric star formation history, a two-component dust attenuation model with a flexible attenuation curve, variable stellar metallicity, and a flexible dust emission template powered via energy balance. Nebular line and continuum emission is generated self-consistently through use of CLOUDY \citep{ferland13} model grids from \citet{byler16}. All reported parameter estimates are the median of the marginalized posterior probability function, with the 16th and 84th percentiles reported as $1\sigma$ error bars. In addition to the AGN templates described in the following section, we make several updates to the \citet{leja17} model:
\begin{itemize}
\item We rewrite the SFH prior in terms of a series of transformations suggested by \citet{betancourt12}. These transformations remove the inefficiency of sampling a Dirichlet distribution with an MCMC algorithm by guaranteeing that every MCMC draw is in the Dirichlet support. This has no effect on the \citet{leja17} physical model: the SFH prior in each bin remains centered on sSFR$_{\mathrm{bin}}$ = 1/t$_{\mathrm{univ}}$.
\item We implement an MCMC convergence criteria based on the Kullback-Leibler divergence of the marginalized PDFs for each model parameter. This quantitative convergence criteria is helpful in both ensuring chain convergence and in maximizing the effective sample size from a given MCMC chain, producing results which are more robust to the numerical noise inherent in MCMC fits.
\item We implement an informative prior on the ratio of the birth-cloud dust to the diffuse dust in the \citet{charlot00} dust model: a Gaussian centered on $\tau_{\mathrm{birth}}/\tau_{\mathrm{diffuse}}$ = 1 with a width of $\sigma$=0.3. This is implemented to ensure consistency with indirect measurements of this ratio (e.g., \citealt{calzetti94,price14}).
\item We broaden the priors on the \citet{draine07} IR template parameters so that they can describe the hot, dusty star-forming galaxies this study focuses on. The new priors are flat over $0 < $\gammae{}$<1$, $0.1 < $\umin{}$<25$, and $0<$\qpah{}$<10$.
\end{itemize}

\subsection{CLUMPY AGN Templates}
\label{sec:agntemplates}
We adopt AGN templates from the \citet{nenkova08a,nenkova08b} CLUMPY models. The CLUMPY AGN models are created by shining an AGN incident broken power-law spectrum through a clumpy dust torus medium using radiative transfer approximations. The analytic formalism describing radiative transfer through a clumpy dust medium is laid out in \citet{nenkova08a}, while the full AGN torus emission model is in \citet{nenkova08b}. Clumpy AGN torus models have been successful in explaining the observed MIR characteristics of AGN in the nearby universe (e.g., \citealt{mor09,honig10}).
\begin{figure*}[ht!]
\begin{center}
\includegraphics[width=0.95\linewidth]{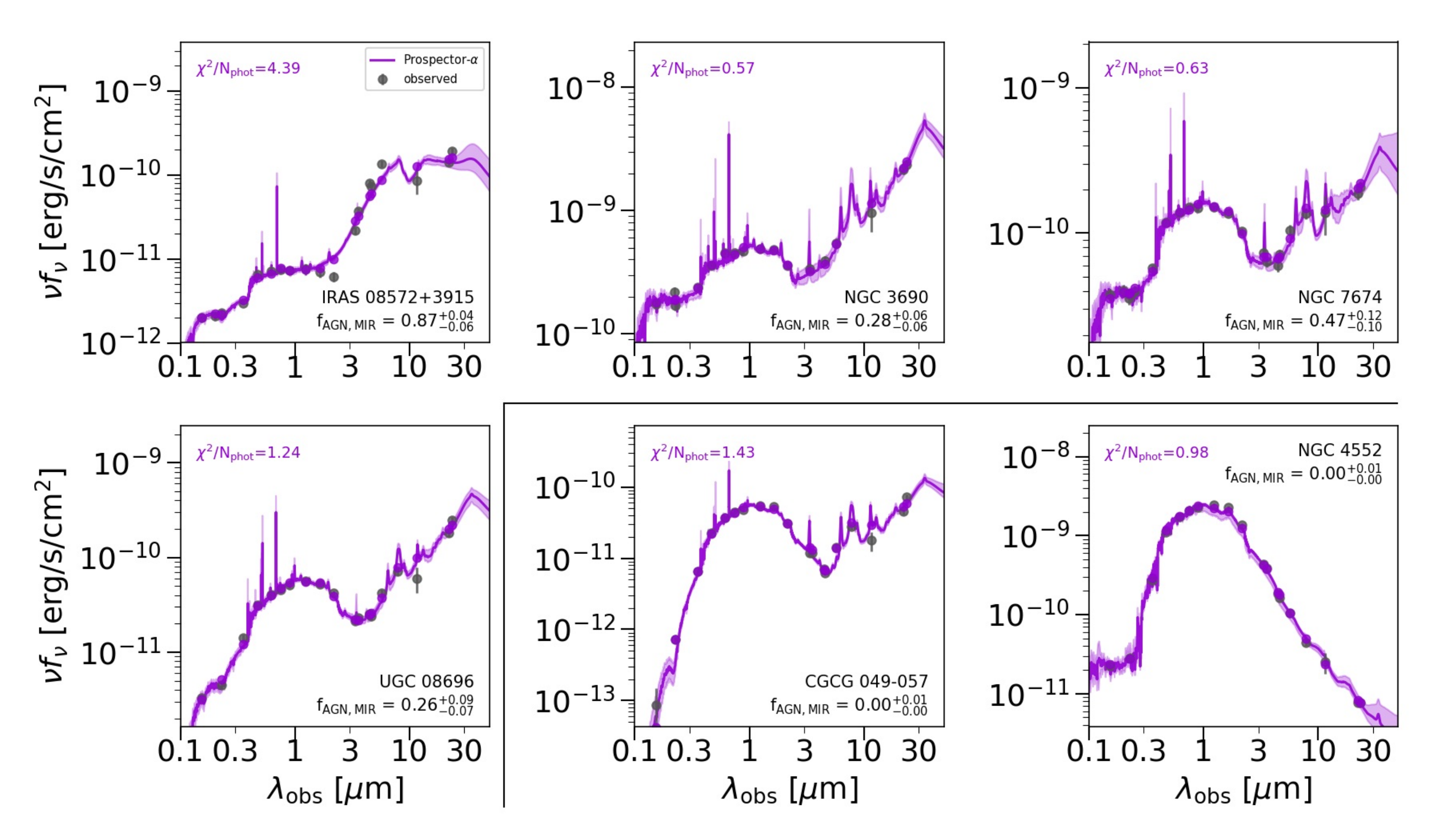}
\caption{Best-fit \mname{} photometry (purple) to the observed broadband UV-IR SEDs of six galaxies (grey circles). The median spectrum from the posterior is also shown, with the 16th and 84th percentiles as light purple regions. Four of the galaxies have significant \fmir{} from the photometric fits, while the two in the corner box are typical star-forming (NGC 7674) and quiescent (NGC 4552) galaxies with no significant AGN contribution. The photometry in all cases is well-fit by the \mname{} model.}
\label{fig:phot_agn_only}
\end{center}
\end{figure*}

The CLUMPY AGN templates are incorporated in the FSPS \citep{conroy09b} source code, with the following assumptions:
\begin{itemize}
\item The ratio of the outer to inner radii, a measure of torus thickness, is fixed as $Y = R_0 / R_d = 10$.
\item An average of $N_0$ = 10 clumps along a radial equatorial ray.
\item A radial density power law exponent of $q=2.0$.
\item A Gaussian angular distribution, of width $\sigma=45^{\circ}$.
\item A standard ISM dust composition, following \citet{ossenkopf92} and \citet{draine03}.
\item A fixed viewing angle of 40$^\circ$.
\item The native AGN torus emission is attenuated by the diffuse dust within the galaxy.
\end{itemize}
Full descriptions of these parameters are available in \citet{nenkova08a,nenkova08b}. Only torus dust emission is significant in this model; the UV and optical emission from the central engine is largely obscured by the AGN dust torus, and the UV/optical emission which does leak out is then attenuated by the galaxy dust attenuation model. We do not model emission directly from the accretion disk, which takes the form of a power law (f$_{\nu} \propto \nu^{-\alpha}$) in the optical and UV. While a strong power-law contribution to a galaxy SED will result in substantial bias in the resulting galaxy properties \citep{cardoso17}, a UV-optical power-law is degenerate with stellar emission: mock tests suggest that up to 26\% of the luminosity at 4000 \AA{} can come from an AGN accretion disk without being identified by an SED-fitting routine \citep{cardoso17}. Given their rarity and their degeneracy with the stellar contribution, it may be best to identify emission from naked accretion disks with other methods, such as optical emission line flux ratios or morphology, X-ray flux, or optical morphology \citep{juneau13}.
\begin{figure*}[ht!]
\begin{center}
\includegraphics[width=0.95\linewidth]{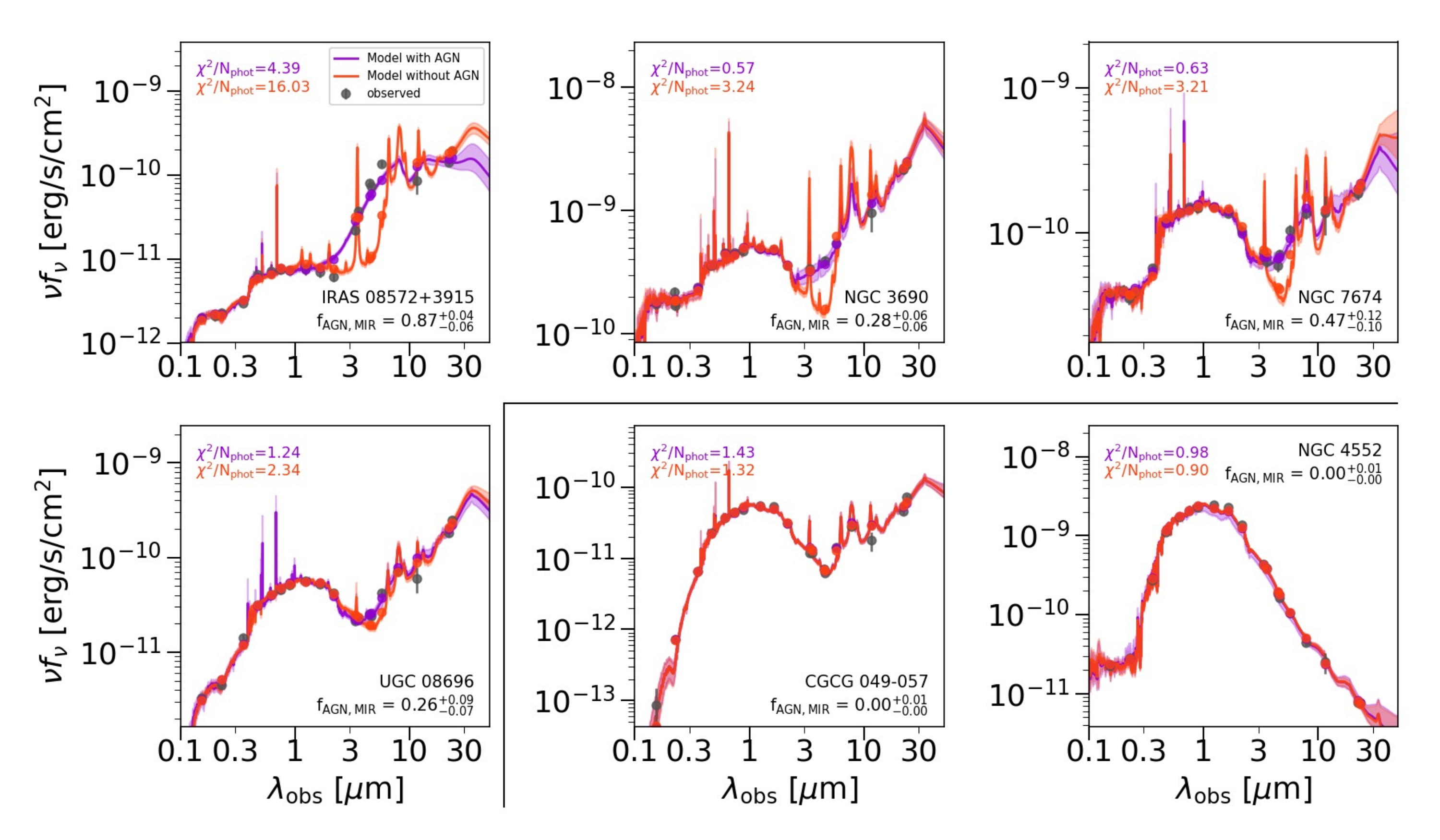}
\caption{Same as Figure \ref{fig:phot_agn_only}, but now also showing the \mname{} model without the AGN templates included in the fit. For 4 of displayed galaxies (10\% of the full sample) the quality of the model fit in the MIR is substantially improved by AGN parameters. The model with AGN also makes different predictions for both line strengths and the continuum shape. Most galaxies in the sample show no significant change in $\chi^2$, similar to the SEDs in the corner box in the lower right (the slight difference in $\chi^2$ for these galaxies is due to numerical noise in the best-fit parameters from an MCMC search).}
\label{fig:phot_both}
\end{center}
\end{figure*}

The CLUMPY AGN models within FSPS are characterized by two variables: \fagn{}, the fraction of bolometric luminosity which comes from the AGN, and \tauagn{}, the optical depth of an individual dust clump at 5500 \AA{}. In this work we reparameterize \fagn{} into the more physically-motivated \fmir{}, defined as the fraction of the total 4-20$\mu$m luminosity which comes from the AGN. The effect of these two parameters on the SED of a typical star-forming galaxy is shown in Figure \ref{fig:sedresponse}. The CLUMPY models provide SEDs for discrete values of \tauagn{}. Linear interpolation between these discrete models is used to create a smooth parameter space.

In \prospector{}, a log-uniform prior is adopted for \fagn{}, with an allowed range of $1\times10^{-5} < $ \fagn{} $ < 3$. A log-uniform prior describes the observed power-law distribution of black hole accretion rates (e.g., \citealt{aird17,georgakakis17}). We note that while \fagn{} is likely to be correlated with accretion rate, the exact relationship is complicated by both the nonlinear correlation between AGN bolometric luminosity and accretion rate \citep{shakura73} and the fact that the fraction of AGN bolometric luminosity re-processed into MIR emission by the surrounding environment is likely highly variable \citep{urry95}. We further note that the adopted lower limit on \fagn{} is below the detectability limit for the \citet{brown14} photometric coverage and S/N. As \fagn{} is generally well-constrained by the data, it is good practice to have priors somewhat wider than the typical physical range of the parameter: this ensures that the posteriors are data-driven rather than prior-driven, and makes it simple to identify catastrophic failures. The adopted lower limit has no effect on the science results.

A log-uniform prior on \tauagn{} is adopted between $5 <$ \tauagn{} $<150$, as the SED response to logarithmic changes in \tauagn{} is approximately linear (see Figure \ref{fig:sedresponse}).

This work focuses on ULIRGs and starbursts, which often display a deep 10$\mu$m silicate absorption feature \citep{spoon07}. This silicate feature is not present in the CLUMPY models, as these deep silicate absorption features require the nuclear source to be embedded in a smooth, optically-thick dust geometry \citep{levenson06}, whereas the \citet{nenkova08b} dust torus models are clumpy and produce shallow absorption features \citep{nenkova02}. Accordingly, the region around the 10$\mu$m silicate feature is masked in this work. The 18$\mu$m silicate feature has many of the same geometric dependencies \citep{sirocky08}. However, as it is a much weaker absorption feature, we do not mask it.

The Prospector parameter files used in this work are available online, both for models with\footnote{\url{https://github.com/jrleja/prospector_alpha/blob/master/parameter_files/brownseds_agn/brownseds_agn_params.py}} and without\footnote{\url{https://github.com/jrleja/prospector_alpha/blob/master/parameter_files/brownseds_np/brownseds_np_params.py}} AGN. The Github commit hash is \texttt{af642bdee7e3de04739 191ea9fd446be218975c5}.

\section{Fitting the photometry and predicting the spectra}
\label{sec:results}
\subsection{SED fits to the broadband photometry}
\label{sec:resid}
\begin{figure*}[ht!]
\begin{center}
\includegraphics[width=0.85\linewidth]{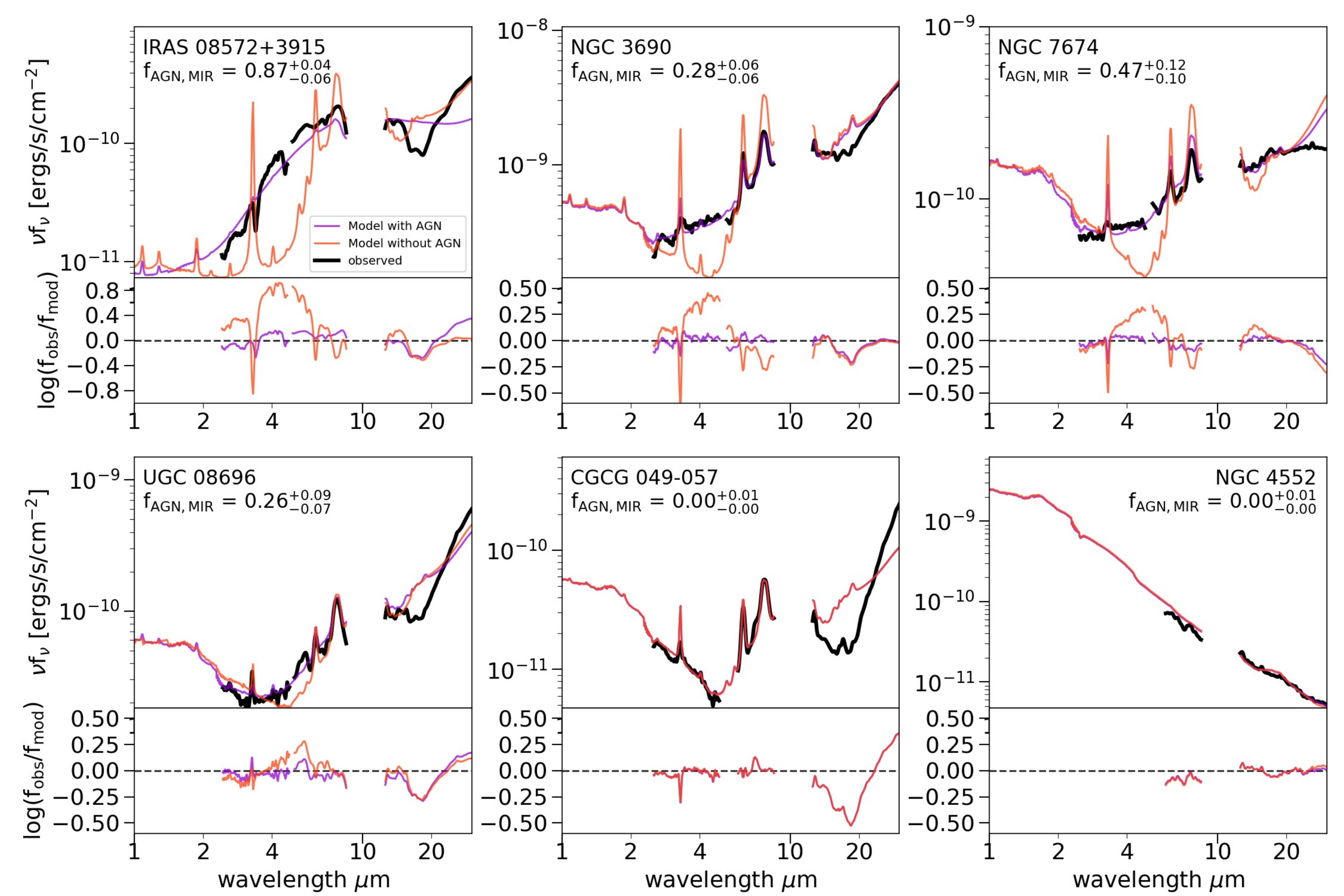}
\caption{Model predictions of the MIR spectra compared to observed {\it Akari} grism and \spitzer{}/IRS spectra (thick black lines), both for models with and without AGN (purple and red lines). The model is fit only to the broadband photometry, not the spectra. Both the hot MIR continuum and PAH line strengths are better predicted when an AGN component is included. The 10$\mu$m absorption feature is masked for reasons discussed in Section \ref{sec:agntemplates}.}
\label{fig:individual_resid}
\end{center}
\end{figure*}
We fit the galaxies in the \citet{brown14} sample with full \mname{} model, including AGN. Representative fits to the broadband photometry and the model spectra associated with those fits are shown in Figure \ref{fig:phot_agn_only}. Four of the highlighted galaxies have significant AGN contribution to the mid-infrared photometry, as measured by \fmir{}, while two of them have no significant AGN contribution. The model provides an excellent fit to the far-UV to MIR photometry for these galaxies as indicated by the small residuals for most of the fits.

The fit to IRAS 08572+3915 is somewhat lower quality than the others, with $\chi^2$/N$_{\mathrm{phot}}$ = 4.39. It is the most powerful AGN in our sample, with $\sim$90\% of its MIR luminosity coming from the AGN. More detailed modeling finds that 90\% of the {\it bolometric} luminosity comes from the AGN, and this galaxy is likely the most luminous infrared galaxy in the $z<0.2$ universe \citep{efstathiou14}. It is remarkable that despite the unusual nature of IRAS 08572+3915, it is still fit reasonably well by the simple two-parameter AGN model introduced in Section \ref{sec:agntemplates}. 

In Figure \ref{fig:phot_both}, we compare these fits to fits from a model without AGN. For the four galaxies with nonzero values of \fmir{}, the models without AGN provide considerably worse fits to the photometry than those with AGN. These galaxies are representative of the $\sim$10\% of the sample which is significantly improved by the AGN model. For these galaxies, there is a notable difference in the shape of the IR SED, and in the strength of the emission and absorption lines. These differences are linked to differences in the underlying physical model, which are explored in detail in Section \ref{sec:galeffect}.

For the two galaxies with \fmir{} $\sim0$, the fit is identical to a model without AGN. The majority of the sample is unaffected by the addition of an AGN parameter.

\subsection{Improved predictions for the spectra}
\label{sec:specpred}
\begin{figure*}[ht!]
\begin{center}
\includegraphics[width=0.95\linewidth]{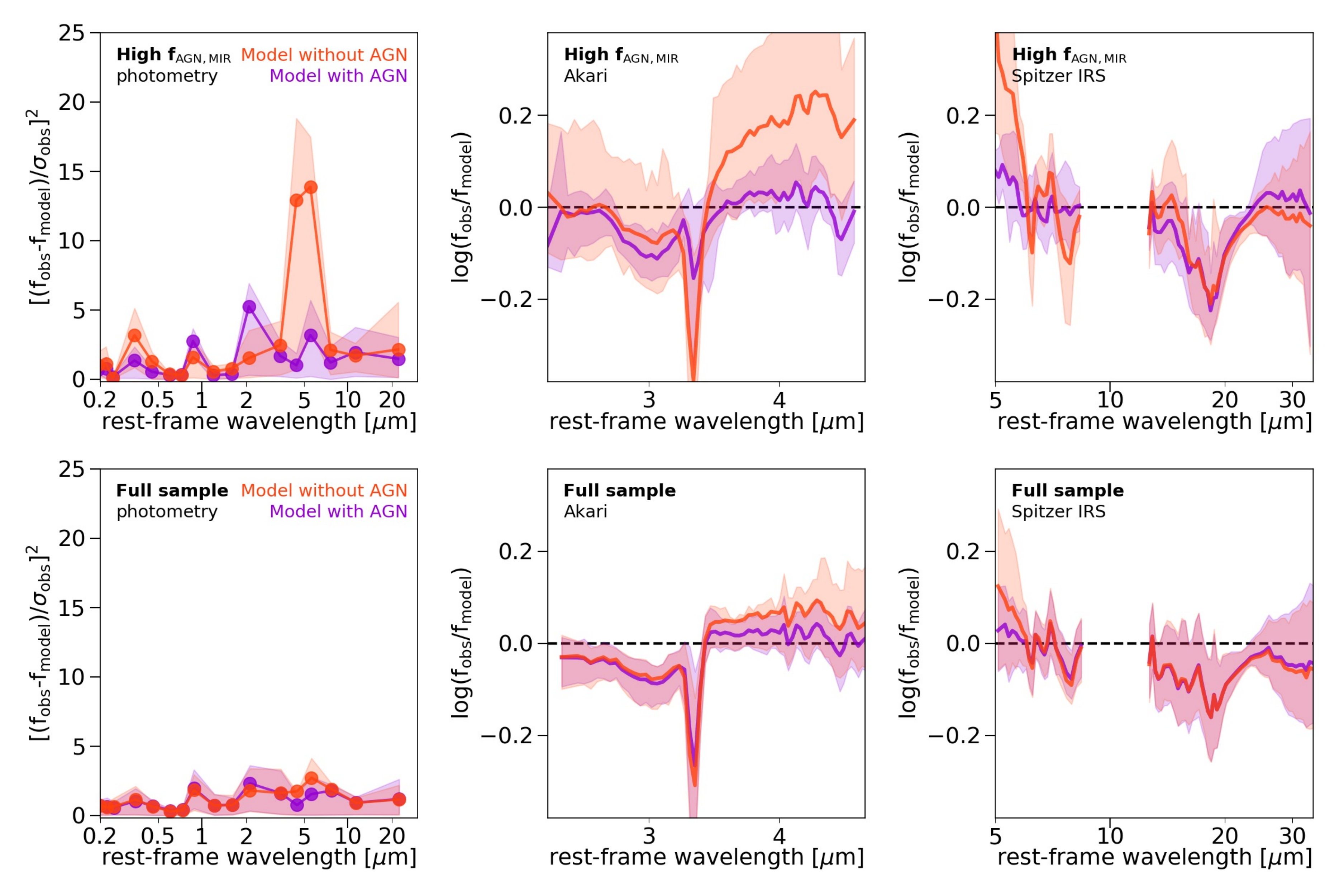}
\caption{Spectral and photometric residuals for the \citet{brown14} sample. The top panels show only galaxies with \fmir{} $> 0.1$, while the bottom panels show the whole sample. The residuals improve when AGN are allowed in the fit, both in the MIR photometry and in the spectra between 3-9$\mu$m. The upper left panel shows the average photometric $\chi^2$ (solid lines) with the 84th and 16th sample percentiles as transparent colors. The other panels show the average spectroscopic residuals, with the 1$\sigma$ population variance shown in transparent colors. The spectra are not fit, but instead predicted from the model fit to the photometry. The 10$\mu$m absorption feature is masked for clarity.}
\label{fig:resid}
\end{center}
\end{figure*}
Here we compare model spectra, predicted solely from the fits to the photometry, to the observed {\it Akari} and \spitzer{} spectra. Figure \ref{fig:individual_resid} shows predictions from SED fits with and without AGN. The differences between the two models range from the removal of a factor of 10 residual in the MIR for IRAS-08572+3915, to no change for galaxies with \fmir{} $\sim 0$. Improvement is seen in both the predicted continuum shape and in the predicted PAH line strengths. The smooth, relatively featureless power-law predicted by the AGN component matches very well to the observed 3-8$\mu$m spectra of galaxies with AGN components. The galaxies without AGN components show very little change in the predicted MIR spectra, as expected.
\begin{figure*}[t!h]
\begin{center}
\includegraphics[width=0.979\linewidth]{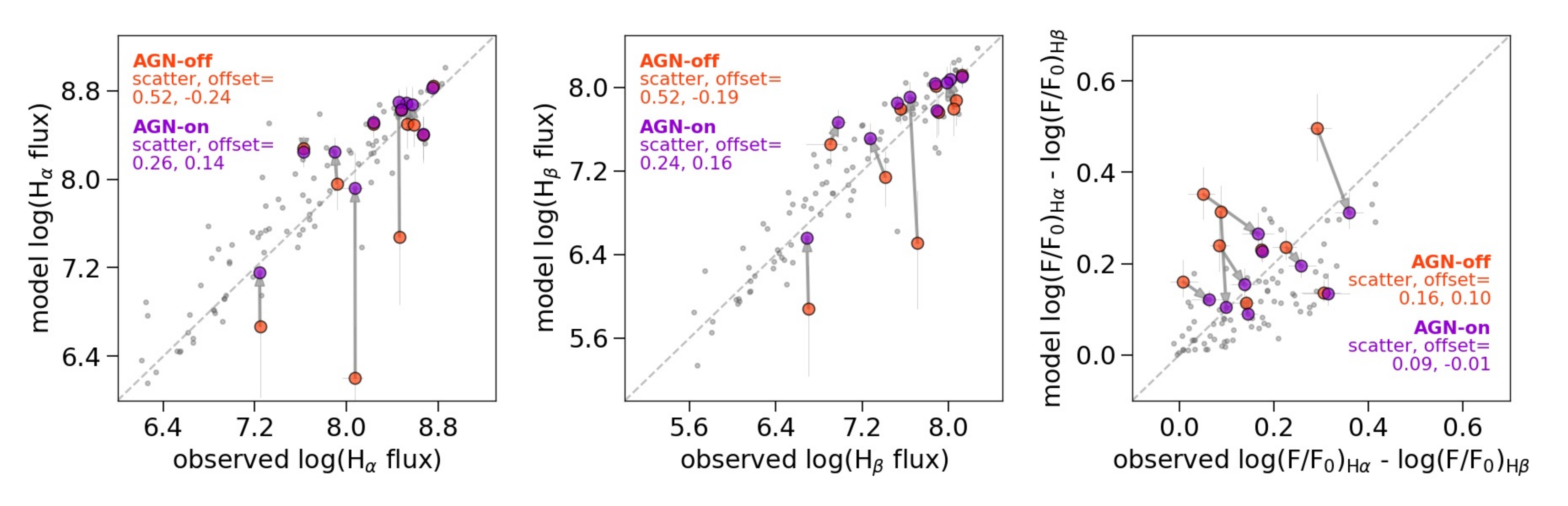}
\caption{Observed \halpha{}, \hbeta{} line luminosities in units of L$_{\odot}$ and observed Balmer decrements compared to SED model predictions from \mname{} fits to the photometry. AGN templates decrease both the scatter and the offset in comparison to the observables. Galaxies with strong AGN are colored circles, while the main galaxy sample is in light grey circles. The change in each property as AGN templates are turned on is indicated by a grey arrow. The mean offset and biweight scatter \citep{beers90} between model and observed properties for galaxies with strong AGN is labeled in each panel. Only points with an observed error of $<0.1$ dex are shown.}
\label{fig:delta_specpars}
\end{center}
\end{figure*}

Figure \ref{fig:resid} shows the spectral and photometric residuals for the \citet{brown14} sample after fitting the photometry with the \mname{} model. The inclusion of an AGN template improves the residuals for the entire sample, in addition to the case-by-case improvement already highlighted. Notably, after adding the AGN template, the residuals in galaxies with high \fmir{} closely resemble the residuals of the full sample, suggesting that the AGN component is well-modeled. The decrease in photometric residuals comes primarily in the MIR regime. The MIR spectral residuals improve by an average $\sim0.1$ dex at 3-9$\mu$m. Outside of the MIR wavelengths, both the spectroscopic and photometric residuals are largely insensitive to the presence of an AGN template.

We also investigate the change in model predictions for the observed \halpha{} and \hbeta{} luminosities when the AGN template is included. Figure \ref{fig:delta_specpars} show the observed \halpha{} and \hbeta{} emission line luminosities and their ratio (the Balmer decrement), compared to the \mname{} predictions for these properties. The predictions are based solely on fits to the broadband photometry. We show changes in these predictions for galaxies with \fmir{} $> 0.1$, while the rest of the sample is shown as light grey circles.

The AGN templates remove many of the outliers in these comparisons, returning them to the 1:1 relationship. This improvement is quantified by the change in mean offset and biweight scatter in each panel. In each case, the mean offset and biweight scatter decrease when AGN templates are turned on. Physically, \halpha{} and \hbeta{} luminosities are sensitive to the recent star formation rate, the dust attenuation model, and the stellar metallicity \citep{leja17}, while the Balmer decrement directly measures the reddening due to dust \citep{osterbrock89}. This implies that the accuracy of the dust attenuation model and the model star formation rates is improved by including AGN templates.

Interestingly, not only do the model predictions change when AGN parameters are added, but the observed \halpha{} and \hbeta{} luminosities also change. This change is the most dramatic for the observed Balmer decrement. This change occurs because a stellar continuum sampled from the model is used to correct the observed \halpha{} and \hbeta{} luminosities for the underlying stellar absorption. The change in the underlying stellar absorption can be substantial: without AGN templates, many of the galaxies hosting AGN have a large mass fraction in 0.1-1 Gyr-old stars (as discussed in Section \ref{sec:sfh}), which is where the depth of the Balmer absorption lines is maximized (e.g., \citealt{bezanson13}). Thus, the AGN templates are important not just for predicting emission line strengths, but also for actually measuring the Balmer emission line luminosity from the observed spectrum.

\section{The effect of an AGN component on derived galaxy properties}
\label{sec:broader_galeffect}
\begin{figure*}[th!]
\begin{center}
\includegraphics[width=0.95\linewidth]{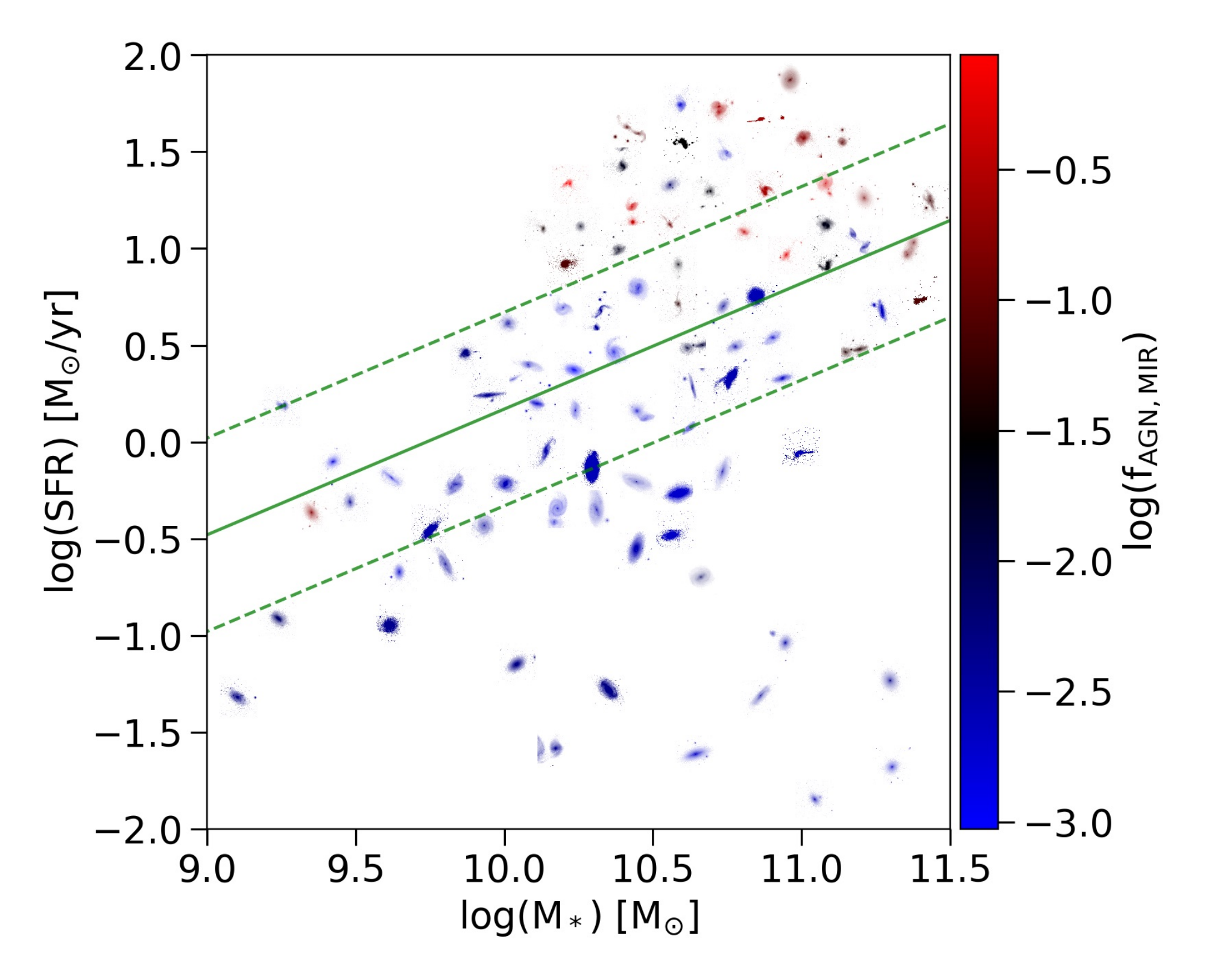}
\caption{Galaxies from the \citet{brown14} catalog shown on the star-forming sequence. Out of the 96 galaxies in the \citet{brown14} catalog which lie in the bounds of this figure, 89 are shown, due to space constraints. Galaxies are color-coded by their \fmir{} from \mname{}. The images are SDSS $i$-band. The green line is the star-forming sequence from \citet{salim07}, and the dashed lines indicate the measured dispersion in this sequence (0.5 dex). Above 10$^{10}$ M$_{\odot}$, galaxies hosting AGN tend to live above the star-forming sequence.}
\label{fig:sf_seq}
\end{center}
\end{figure*}
Here we show the distribution of AGN in the galaxy population according to the \mname{} model, and then explore how their derived physical properties change when an AGN component is included.  The analysis focuses on galaxies with a strong AGN contribution, defined as \fmir{} $> 0.1$ from the fits with \mname{}. This criteria includes 13 out of 129 galaxies in the sample.

Section \ref{sec:props} describes the properties of galaxies with a significant AGN component, while Section \ref{sec:galeffect} describes how the derived galaxy properties change when AGN templates are excluded from the fit.

\subsection{Properties of galaxies with significant AGN components}
\label{sec:props}
Figure \ref{fig:sf_seq} shows randomly selected galaxies from the \citet{brown14} sample on the star-forming sequence. Each galaxy is represented by its SDSS $i$-band image, taken directly from the \citet{brown14} online repository, and colored according to \fmir{}. The stellar masses and star formation rates are taken from the \mname{} model. It is clear from this figure that most galaxies with significant \fmir{} are massive galaxies which live on or above the star-forming sequence, and often show irregular morphologies. We emphasize that the \citet{brown14} sample is not an unbiased sample of the entire galaxy population, and so extrapolations of MIR AGN characteristics to the general population should be performed with caution. Furthermore, the location of AGN within the star-forming sequence likely depends on the AGN selection technique: for example, \citet{leslie16} select AGN by their optical emission line ratios and find that they typically are located below the star-forming sequence. Nonetheless, it is clear that in these galaxies, star formation and dust-obscured AGN are strongly associated.

\begin{figure*}[ht!]
\begin{center}
\includegraphics[width=0.85\linewidth]{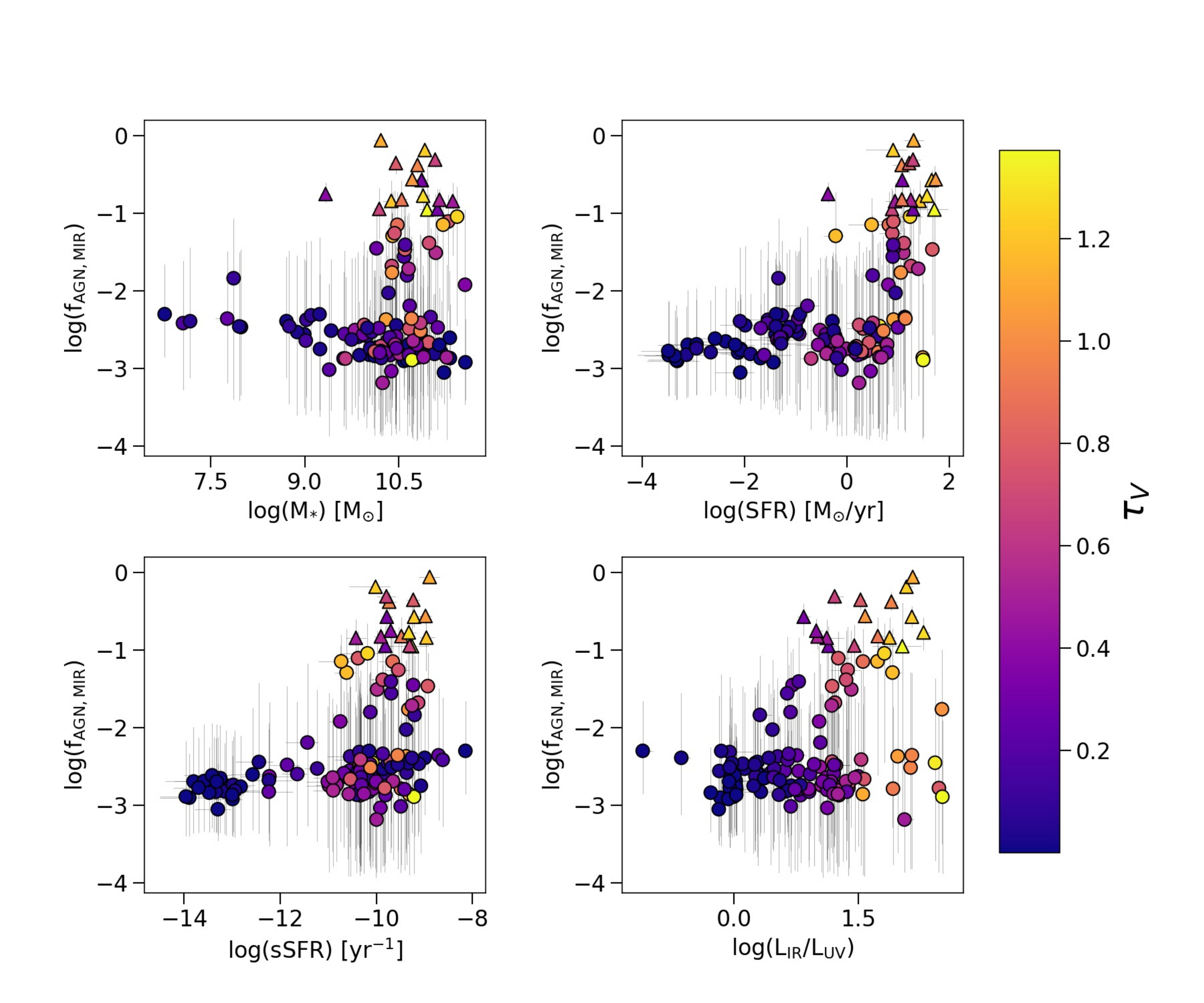}
\caption{Correlation between galaxy properties and the strength of the MIR AGN component derived in \mname{}. MIR AGN live primarily in dusty star-forming galaxies, though not all dusty star-forming galaxies host an AGN. Galaxies with strong AGN (i.e., \fmir{} $>$ 0.1) are triangles, while the rest of the sample is in circles. Points are colored according to their dust optical depth at 5500 \AA{}.}
\label{fig:galaxprop}
\end{center}
\end{figure*}

Figure \ref{fig:galaxprop} takes a closer look at the properties of MIR AGN hosts by plotting the correlation between \fmir{} and other galaxy properties. In particular, the behavior of \fmir{} with log(\lir{}/\luv{}) is striking. There are two distinct sequences in this panel, one which is flat in \fmir{} with log(\lir{}/\luv{}) and one which shows a linear relationship between \fmir{} and log(\lir{}/\luv{}). This suggests that a bifurcation in AGN activity at high log(\lir{}/\luv{}), perhaps related to different modes of star formation. However, the unusual selection function of the \citet{brown14} sample stymies a detailed exploration of this correlation (or indeed, may cause the correlation).

More general, Figure \ref{fig:galaxprop} suggests that AGN in the sample live in log(M/M$_{\odot})\sim 10.5$ galaxies, with specific star formation rates (sSFR) $\sim 10^{-10}$ yr$^{-1}$. MIR AGN host galaxies tend to be dustier than the average galaxy, though they span a range of dust attenuations, all the way down to modest optical depths of 0.3 at 5500 \AA{}. In summary, MIR AGN tend to live in dusty star-forming galaxies, though not all dusty star-forming galaxies host a MIR AGN.

\subsection{Effect of AGN templates on galaxy properties}
\label{sec:galeffect}
\begin{figure*}[t!h]
\begin{center} 
\includegraphics[width=0.979\linewidth]{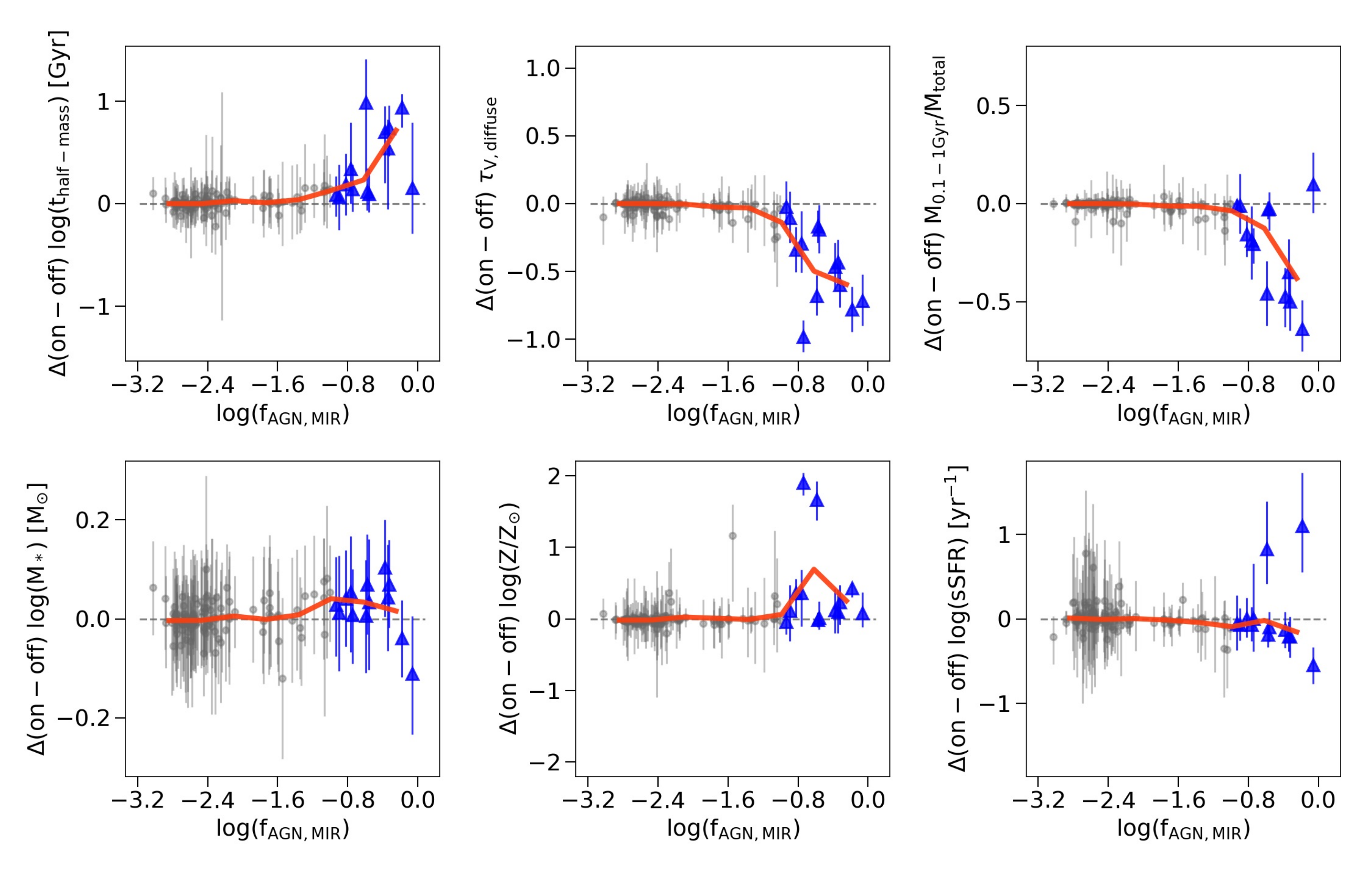}
\caption{Change in galaxy properties as a function of \fmir{} between fits with and without AGN. The running average is shown as a thick red line. On average, stellar ages increase, dust attenuation increases, the mass fraction of stars aged 0.1-1 Gyr decreases substantially (emission from circumstellar AGB dust, which peaks at $\sim$0.1-1 Gyr, can mimic the hot dust emission from AGN), stellar masses increase slightly, stellar metallicities increase, and sSFRs do not change (though with considerable scatter at high \fmir{}).}
\label{fig:delta_galpars}
\end{center}
\end{figure*}

Figure \ref{fig:delta_galpars} shows how key galaxy properties change as the AGN model is turned on. These changes are shown as a function of \fmir{}. An increasing \fmir{} increases the stellar age by up to 1 dex, decreases the  dust optical depth, strongly decreases the fraction of mass formed in the 0.1-1 Gyr time, slightly increases stellar mass, increases the stellar metallicity, and has a strong though variable effect on specific star formation rate.

These changes in galaxy properties stem from the SED model being largely unable to model an excess of emission at 3-8$\mu$m (see Figure \ref{sec:resid}) without an AGN template. The \mname{} model has two methods to create large amounts of 3-8$\mu$m emission with stars, dust, and gas alone: either increasing the amount of energy attenuated and re-radiated by galactic dust by altering the SFR, stellar metallicity, and dust attenuation properties, or by contributing directly to the 3-8$\mu$m luminosity via emission from hot AGB circumstellar dust by increasing the fraction of stars aged 0.1-1 Gyr. These are discussed separately.

\subsubsection{Dust attenuation, stellar metallicity, and star formation rates}
The spectral signature of hot dust emission can be mimicked without an AGN template by instead increasing the energy absorbed and re-radiated by dust in the host galaxy. Due to the degeneracy between dust, metallicity, and SFH in the broadband UV-NIR SED \citep{bell01}, it is possible to increase the dust emission budget while keeping the UV through NIR SED approximately fixed. Thus, models with and without an AGN component which are fit to the same observed SED will often produce different solutions for SFR, metallicity, and dust attenuation.

\begin{figure}[t!h]
\begin{center}
\includegraphics[width=0.95\linewidth]{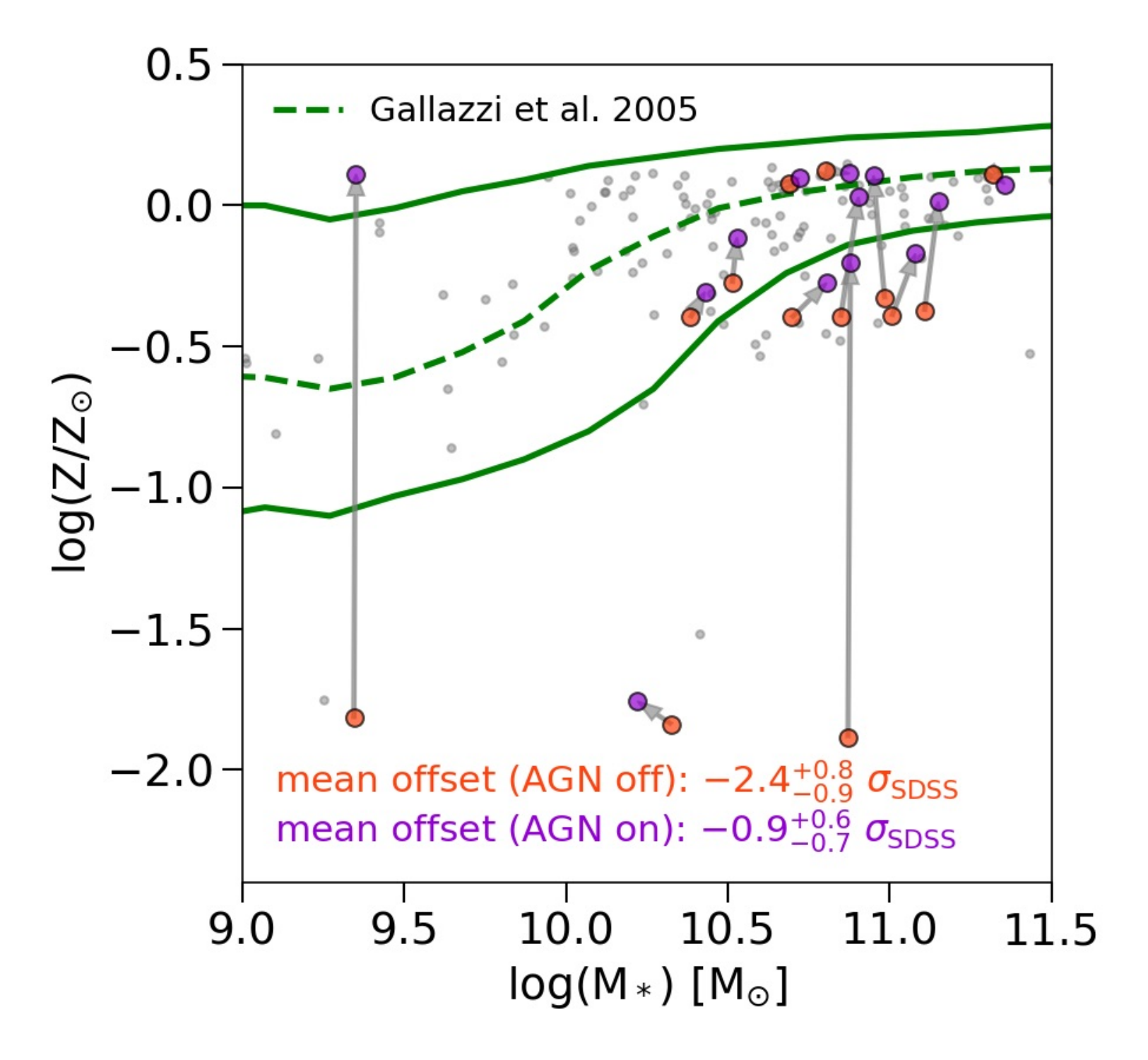}
\caption{AGN host galaxies show much better agreement with the stellar mass--stellar metallicity relationship when fit with an AGN component. The \citet{brown14} sample is shown on the stellar mass--stellar metallicity plane with the SDSS relationship in green. Galaxies with strong AGN are colored circles (red for models with AGN and purple for models without AGN), while the main galaxy sample is shown as light grey circles. Individual galaxies fit by different models are linked with a grey arrow. The lower text denotes the deviation of galaxies hosting strong AGN from the \citet{gallazzi05} relationship, normalized by the width of that relationship. Error bars on this estimate are a combination of errors on the galaxy metallicity estimates and bootstrapping.}
\label{fig:delta_massmet}
\end{center}
\end{figure}

We can assess the accuracy of the galaxy parameters derived with and without AGN templates by comparing to independently measured spectral features which are sensitive to these parameters, as in Figure \ref{fig:delta_specpars}. Predictions of the Balmer emission line luminosities and their ratio improve when AGN are included, both in scatter and in offset. This indicates that the dust attenuation and SFR posteriors are significantly more accurate when AGN are included in the model.

In Figure \ref{fig:delta_massmet}, we assess the accuracy of the new stellar metallicities by observing the change of location of galaxies with strong AGN in the stellar mass--stellar metallicity relationship, as compared to the relationship measured by \citet{gallazzi05} in the SDSS. Many of the galaxies hosting AGN were strong outliers in the stellar mass--stellar metallicity relationship when fit with SED models without AGN. On average, they lay 2.3$\sigma$ below the typical galaxy at the same mass. When AGN are added, this bias is substantially decreased, to 0.9$\sigma$, and nearly consistent with no offset within the estimated errors. The error estimates come from a combination of the errors on individual galaxy metallicity estimates, and errors on the population mean, estimate via bootstrapping. 

One potential source of bias is that spectroscopic mass-metallicity relationship is measured only in the central regions of the galaxies (3$\arcsec$ apertures), whereas photometric apertures are on the order of 100$\arcsec$, covering several effective radii. The typical galaxy in the local universe has a gradient of $\sim$-0.1 (dex / effective radius) from MaNGA IFU data \citep{zheng17}, or a gradient across the galaxy of $\sim$0.2 dex as measured statistically \citep{roig15}. These will contribute to the offset of -0.9 dex, but are smaller than the observed effect by a factor of several.

Overall, this analysis suggests that the stellar metallicities obtained with an AGN template are more reliable than those obtained without.

\begin{figure*}[t!h]
\begin{center}
\includegraphics[width=0.95\linewidth]{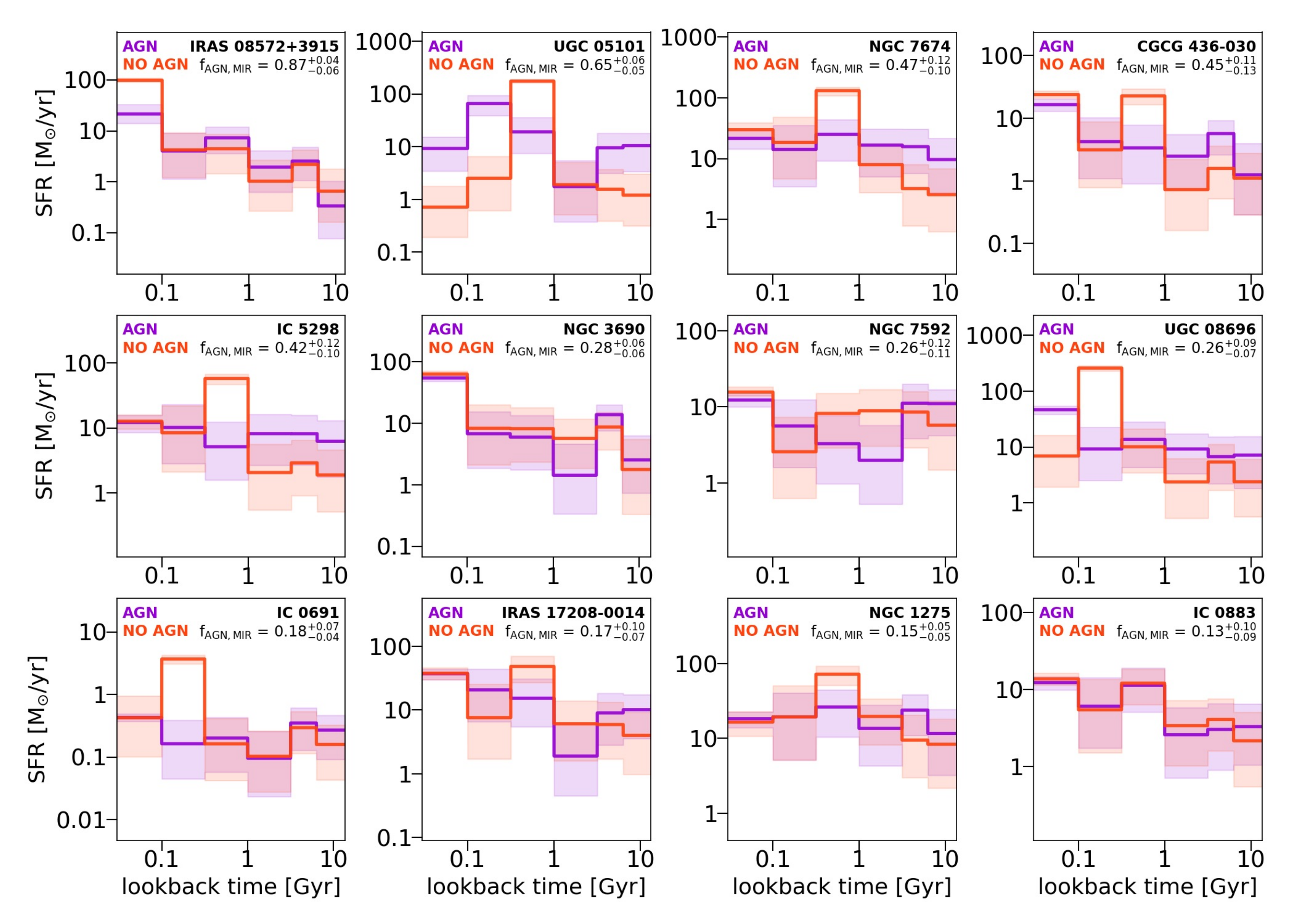}
\caption{Star formation histories from the SED fits, both with and without AGN templates, for the 12 galaxies with the largest \fmir{}. Without AGN templates, galaxies with strong AGN are erroneously assigned large bursts in the 0.1-1 Gyr regime. These bursts are removed when AGN are turned on in the fits. The solid line denotes the median SFR in each time bin, while the shaded region represents the 16th and 84th percentiles of the marginalized posteriors. The galaxies are sorted in order of decreasing \fmir{}.}
\label{fig:sfh_comparison}
\end{center}
\end{figure*}

We note that this is neither an unbiased galaxy sample nor does it necessarily contain a representative sample of AGN hosts, so the location of this sample on the stellar mass--stellar metallicity scaling relationship is not clear. It may lie below the typical scaling relationships. However, given the transitory nature of AGN activity \citep{mushotzky93}, it is unlikely that the current presence of an AGN has a causal effect on galaxy stellar metallicities, which are a product of the steady and inevitable build-up of metals over cosmic time (e.g., \citealt{dave12,derossi17}). We therefore expect at most a weak bias in the stellar metallicities. This expectation is more consistent with results from the AGN-on model.
\subsubsection{Galaxy ages and star formation histories}
\label{sec:sfh}
The second method to generate excess 3-8$\mu$m emission in the \mname{} model is from the emission of hot circumstellar dust around AGB stars. This emission peaks in the mid-infrared \citep{piovan03,kelson10,villaume15} and therefore has a similar spectral signature to AGN-heated dust \citep{alatalo16}. In \mname{}, the contribution of AGB stars to the MIR is maximized in the 0.1-1 Gyr-old star formation history bins. Notably, these bins also contain B and A stars, which have strong optical signatures.

Figure \ref{fig:sfh_comparison} compares the star formation histories for galaxies with strong AGN, fit with and without AGN templates. Without AGN templates, many of the strong AGN galaxies have extremely high star formation rates in the 0.1-1 Gyr SFH bins. The inclusion of AGN templates in the fit smooths out the derived star formation histories and decreases the percent of the total stellar mass in these bins by up to 75\% (Figure \ref{fig:delta_galpars}). This altered SFH has a secondary effect of increasing the mass-weighted age of galaxies by up to a Gyr, as well as slightly increasing their M/L ratios and stellar masses ($\sim$0.05 dex).

In principle, these galaxies could be bona-fide post-starburst galaxies, and the unknown MIR energy source could indeed be a large population of 0.1-1 Gyr-old stars. However, there is evidence against this hypothesis. First and foremost, the AGN-on model prefers AGN over AGB stars. There is not a degeneracy between the two, which is likely due to the fact that an intermediate-age component also manifests itself in the optical. Second, the hot MIR emission comes from the center of the galaxy, rather than following the stellar morphology (Section \ref{sec:wisegradients}). Finally, many of these galaxies are known to have buried AGN which contribute a substantial or dominant fraction of the hot dust emission (Table \ref{table:agnevidence}). The simplest explanation is that the hot dust emission comes from buried AGN, and when this emission is fit without AGN templates, the resulting star formation histories can be highly distorted, resulting in biased age (up to 1 dex) and stellar mass (up to 0.1 dex) estimates.

\section{Correlation between \fmir{} and other AGN Indicators}
\label{sec:agn}
\begin{figure}[ht!]
\begin{center}
\includegraphics[width=0.95\linewidth]{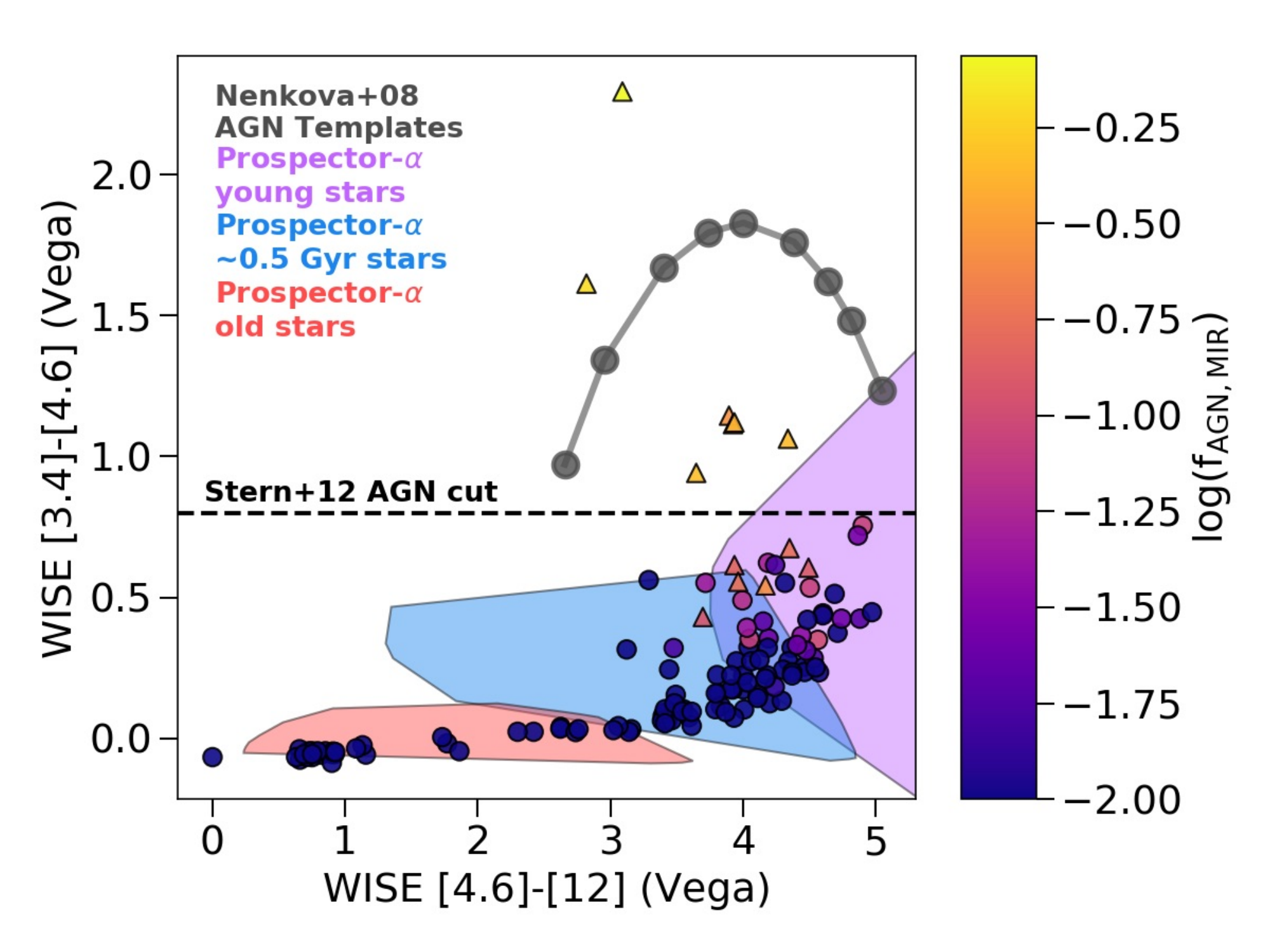}
\caption{\wise{} color-color diagram, colored by \fmir{} from \mname{} fits to the photometry. Galaxies with strong AGN (\fmir{} $> 0.1$) are denoted with triangles, while other galaxies are circles. The colors spanned by the pure \citet{nenkova08b} templates are shown in grey, while the colors spanned by the \mname{} models in three age ranges are shown as colored regions. There is a clear correlation between \fmir{} and red \wise{} [3.4-4.6] colors. Many AGN identified by the \mname{} model are well outside classical color-color AGN selection techniques (e.g. \citealt{stern12}, W1-W2 $>$ 0.8).}
\label{fig:wise_colors}
\end{center}
\end{figure}

Here we investigate whether the high \fmir{} is tracing AGN emission or some alternative source of 3-8$\mu$m energy by comparing to three independent AGN indicators: \wise{} color gradients, X-ray luminosity, and emission line diagnostics. We also compare to classic AGN color-selection techniques, though these techniques are not independent of the method presented here as both rely on integrated photometry.

Finding a clean answer to this question is complicated by the fact that AGN selection techniques at different wavelengths show little overlap. This is understandable, as AGN unification models predict that different signatures of AGN activity demonstrate different viewing angle dependences \citep{urry95}. Multiple studies show that AGN indicators are far from complete, in the sense that they pick out different and often non-overlapping AGN populations \citep{juneau13,trump15}. X-ray identification of AGN is thought to be the most unbiased, as X-rays penetrate through dust and gas very efficiently \citep{mushotzky04}. In practice, however, a large fraction of known AGN are not identified in X-rays:  \citealt{juneau13} find that $\sim$35\% of optical \& IR AGN are undetected in X-rays, while other studies find anywhere between 22\% to 50\% of MIR-selected AGN are detected in X-rays \citep{donley12,cowley16,koss16,ichikawa17}. Other indicators show similar levels of overlap: for example, $\sim$40-50\% of IR-selected AGN have BPT classifications of ``composite" or ``AGN" \citep{goulding09,azadi17}.

Thus, to supplement the AGN evidence assembled in this paper, we also perform a limited literature search for other signatures of AGN activity in these galaxies, such as radio identification, optical emission line morphology, and PAH equivalent width. The full results of this search are shown in Table \ref{table:agnevidence}. 

\subsection{{\it WISE} integrated colors}
In Figure \ref{fig:wise_colors} the observed \wise{} colors for the \citet{brown14} sample are shown, with points colored by \fmir{} from the \mname{} fits. AGN emission is marked by red W1-W2 colors, the result of hot dust emission. A strong but not monotonic relationship exists between the observed W1-W2 colors and \fmir{} from the \mname{} fit.

We also show the MIR AGN photometric selection criteria proposed by \citet{stern12}, which is a simple cut at W1-W2 $>$ 0.8. This technique identifies only 6 MIR AGN in this sample, while the \mname{} model identifies 13 galaxies with \fmir{} $> 0.1$. The comparison with simple color-based AGN selection techniques is discussed further in Section \ref{sec:other_techniques}.

The colors spanned by the \mname{} dust emission models for stars of various ages are also shown in Figure \ref{fig:wise_colors}. These models are generated by creating dust-free stellar populations with constant star formation over 0-100 Myr (young), 100-350 Myr (intermediate), and 8-13.8 Gyr (old) before the time of observation. The MIR colors of these populations are measured over the full range of stellar metallicities. MIR colors are also measured with a dust screen of $\tau_{5500\mathrm{\AA}}$=0.3 added, for the full range of allowed \citet{draine07} dust emission model parameters. The colored regions in Figure \ref{fig:wise_colors} are the convex hull of these model MIR colors. The MIR colors of galaxies with strong AGN are closest to some combination of young and intermediate-aged stars. This suggests that not including AGN in the SED-fitting model will have a significant effect on SFHs, which is shown directly in Section \ref{sec:sfh}.

\subsection{{\it WISE} W1-W2 color gradients}
\label{sec:wisegradients}
\begin{figure*}[t!]
\begin{center}
\includegraphics[width=0.95\linewidth]{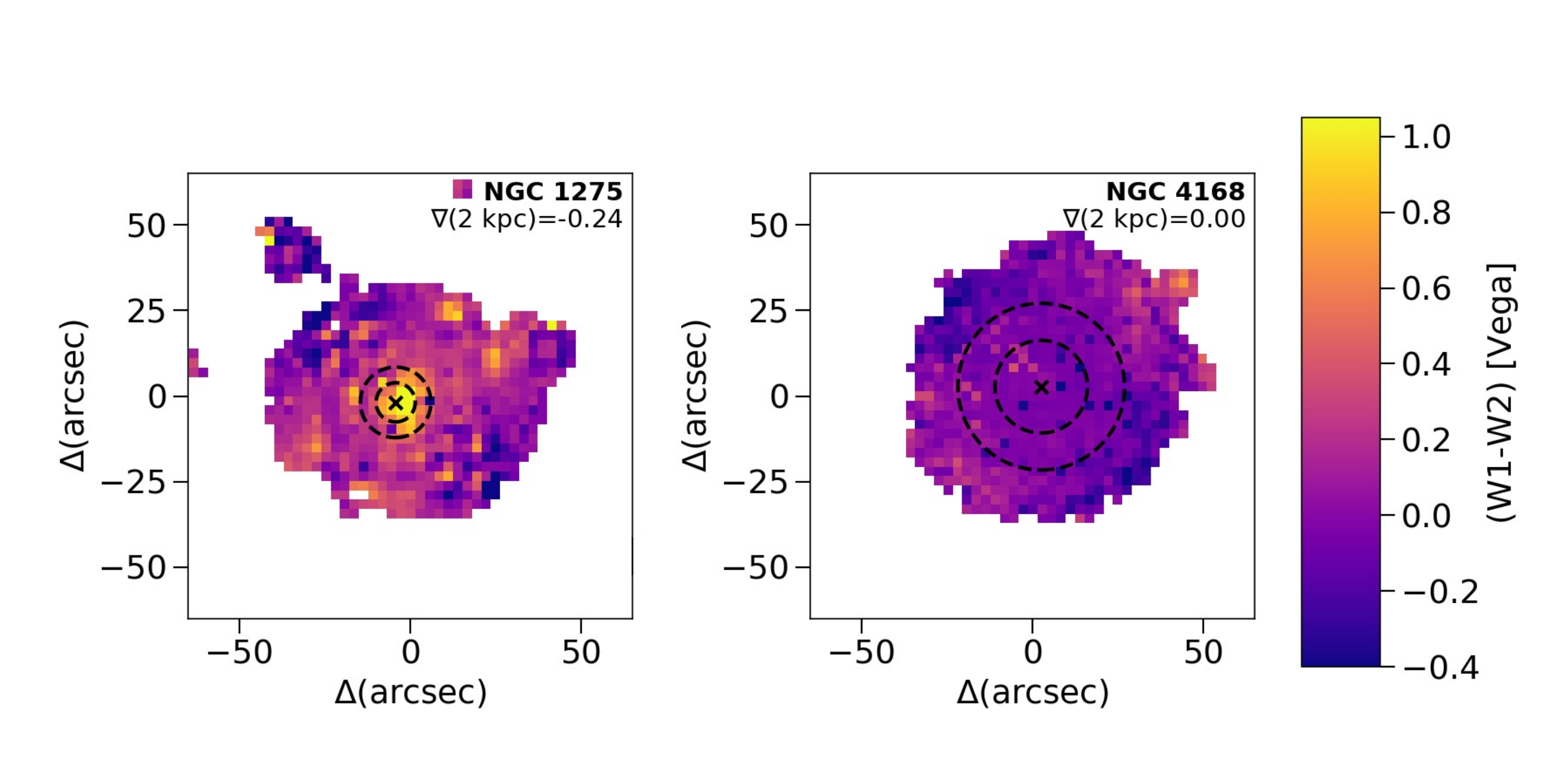}
\caption{Two examples of resolved WISE [3.4]-[4.6] color maps in the \citet{brown14} sample. The two concentric circles mark constant physical distances of 2 and 4 kpc. NGC 1275 has a strong color gradient, while NGC 4168 has no discernible color gradient. The WISE PSF is approximately 6$\arcsec$.}
\label{fig:gradient_example}
\end{center}
\end{figure*}

\begin{figure}[t!]
\begin{center}
\includegraphics[width=0.95\linewidth]{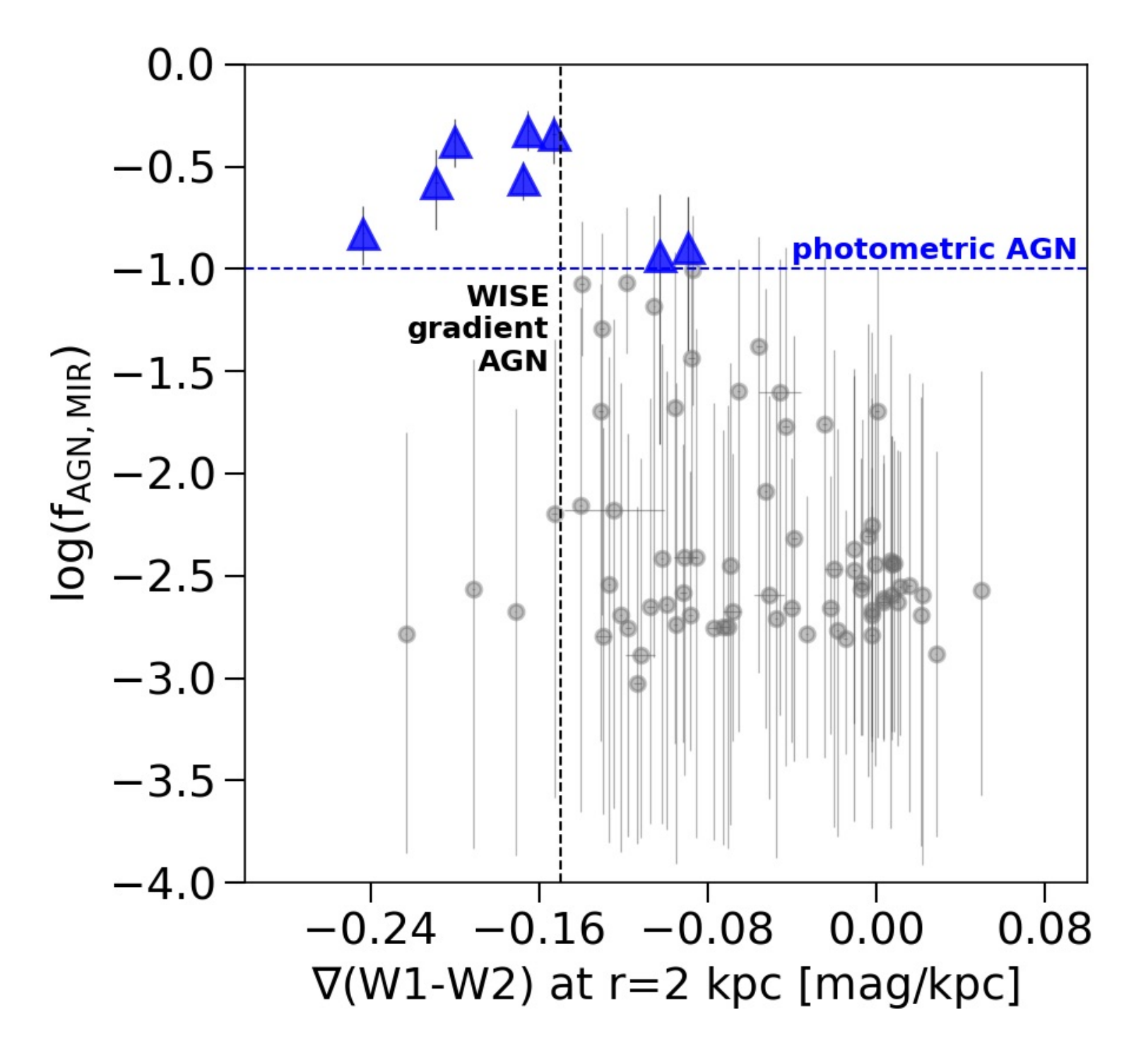}
\caption{\wise{} [3.4]-[4.6] radial color gradients at $r=2$ kpc as a function of \fmir{}. AGN candidates have strong \wise{} color gradients, suggesting their red \wise{} colors are driven by a point source. Strong \mname{} AGN candidates are shown as blue triangles, while other galaxies are shown as grey circles.}
\label{fig:gradient}
\end{center}
\end{figure}
Mid-infrared color maps can be used to shed light on the source of the mid-infrared excess by distinguishing between diffuse and point-source MIR emission. For example, \citet{clemens11} used mid-infrared color maps of Virgo galaxies to show that circumstellar AGB dust rather than AGN emission is responsible for the MIR colors of many Virgo quiescent galaxies. Here, we investigate the morphology of the MIR colors in the \citet{brown14} sample.

Figure \ref{fig:gradient_example} shows two examples of WISE [3.4]-[4.6] color maps, both with and without a strong central color gradient. Figure \ref{fig:gradient} shows the \wise{} W1-W2 radial color gradients, measured at a radial distance of 2 kpc, as a function of \fmir{} for the entire \citet{brown14} sample. Galaxies with low \fmir{} values show a wide range of \wise{} color gradients, from strongly negative to mildly positive. The average \wise{} color gradient is mildly negative, such that the inner regions of each galaxy have redder colors than the outer regions. This is consistent with resolved \herschel{} studies of nearby galaxies in the KINGFISH survey, which were found to have dust temperatures which decrease with radial distance from the center of the galaxy \citep{galametz12}.

The strong AGN sample uniformly have negative \wise{} color gradients, suggesting a redder central source surrounded by a bluer component. Visual inspection of the \wise{} W1 and W2 color and intensity maps for galaxies with strong AGN corroborate this picture, typically showing red centers surrounded by bluer disks. Indeed, several of the strong AGN hosts (NGC 1068, UGC 05101) have very red point-source centers which dominate the light profile, marked by diffraction spikes and rings from the \wise{} PSF.

Conversely, the galaxies with steep W1-W2 radial gradients but low \fmir{} are star-forming galaxies with otherwise typical MIR colors. This may be driven by undetected low-luminosity AGN, or a radial gradient in dust heating related to the typical radial profile of star formation, which declines exponentially \citep{nelson16a}.

The \wise{} color gradients suggest that the mid-infrared colors of galaxies with high \fmir{} are, on average, driven by central point-source engines, likely AGN. We cannot rule out the hypothesis that diffuse hot MIR emission contributes to at least some fraction of the red MIR colors (see Section \ref{sec:hotdust}), but this diffuse emission, if it exists, is not a dominant contributor to the MIR colors of these galaxies.

\subsection{X-ray detections}
\begin{figure}[h!]
\begin{center}
\includegraphics[width=0.95\linewidth]{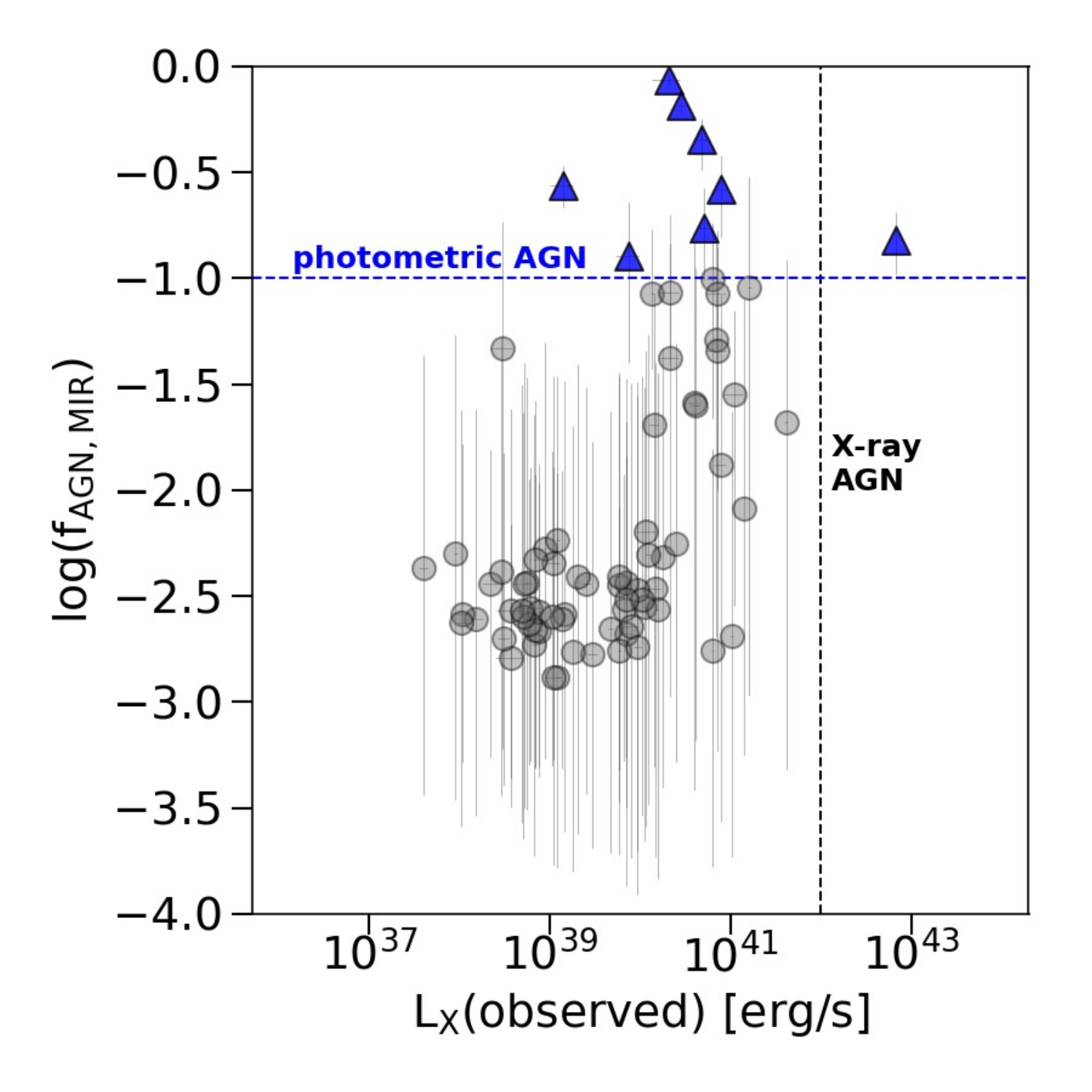}
\caption{Observed X-ray luminosity for galaxies in the sample, as a function of \fmir{}. Galaxies with strong \mname{}-identified AGN are shown as blue triangles, while the rest of the sample is shown in grey. The adopted L$_{\mathrm{X}}$ criteria for identification as an AGN is shown as a dashed line. While most AGN candidates do not meet this criteria, there is still a correlation between \fmir{} and L$_{\mathrm{X}}$. Notably, the sole X-ray detection comes from the AGN candidate with the lowest \tauagn{} in the \mname{} model.}
\label{fig:xray}
\end{center}
\end{figure}
AGN can also be identified through their X-ray emission. X-rays are able to penetrate gas and dust efficiently: at $A_V = 5$ (N$_H \sim 10^{22}$ atoms cm$^{-2}$), the X-ray flux is reduced by only a factor of $\sim$3 \citep{mushotzky04}.

In Figure \ref{fig:xray}, the X-ray luminosity for galaxies with \fmir{} $>0.1$ is shown as a function of \fmir{}. There is a weak relationship between L$_{\mathrm{X}}$ and \fmir{}. Many of the strong AGN candidates from the \mname{} fits are not conclusively identified as AGN by their X-ray emission. Despite this lack of correlation, these are well-known AGN from the literature (see compilation in Table \ref{table:agnevidence}). They primarily live in dusty, highly obscured star-forming galaxies: combined with the lack of strong X-ray fluxes, this suggests that either X-ray flux is partially obscured or Compton-thick, or that these AGN have low accretion rates.

The single X-ray identified AGN is NGC 1275, a Type 1.5 AGN with a high observed X-ray luminosity and the lowest \tauagn{} (\tauagn{} = 10) of the galaxies with strong AGN. This hints that the \tauagn{} derived from MIR broadband photometry may correlate with the X-ray obscuration, though a more complete sample of AGN viewing angles and dust obscuration would be necessary to confirm this correlation.

\subsection{Emission line diagnostics}
The ratios of strong emission lines can be used to classify the main ionization source in a galaxy into one of three categories: young stars, young stars + AGN (``composite``), or AGN \citep{baldwin81,kauffmann03b,kewley06,kewley13a}, known as a BPT diagram. Since the strong emission lines ({\sc [O~iii 5007}]/H$\beta$ and {\sc [N~ii]}/H$\alpha$) are close in wavelength, this AGN diagnostic is robust to dust reddening, though it is sensitive to dilution by star formation and differential extinction in dusty galaxies \citep{mushotzky04,juneau13,trump15}.

\begin{figure}[t!]
\begin{center}
\includegraphics[width=0.95\linewidth]{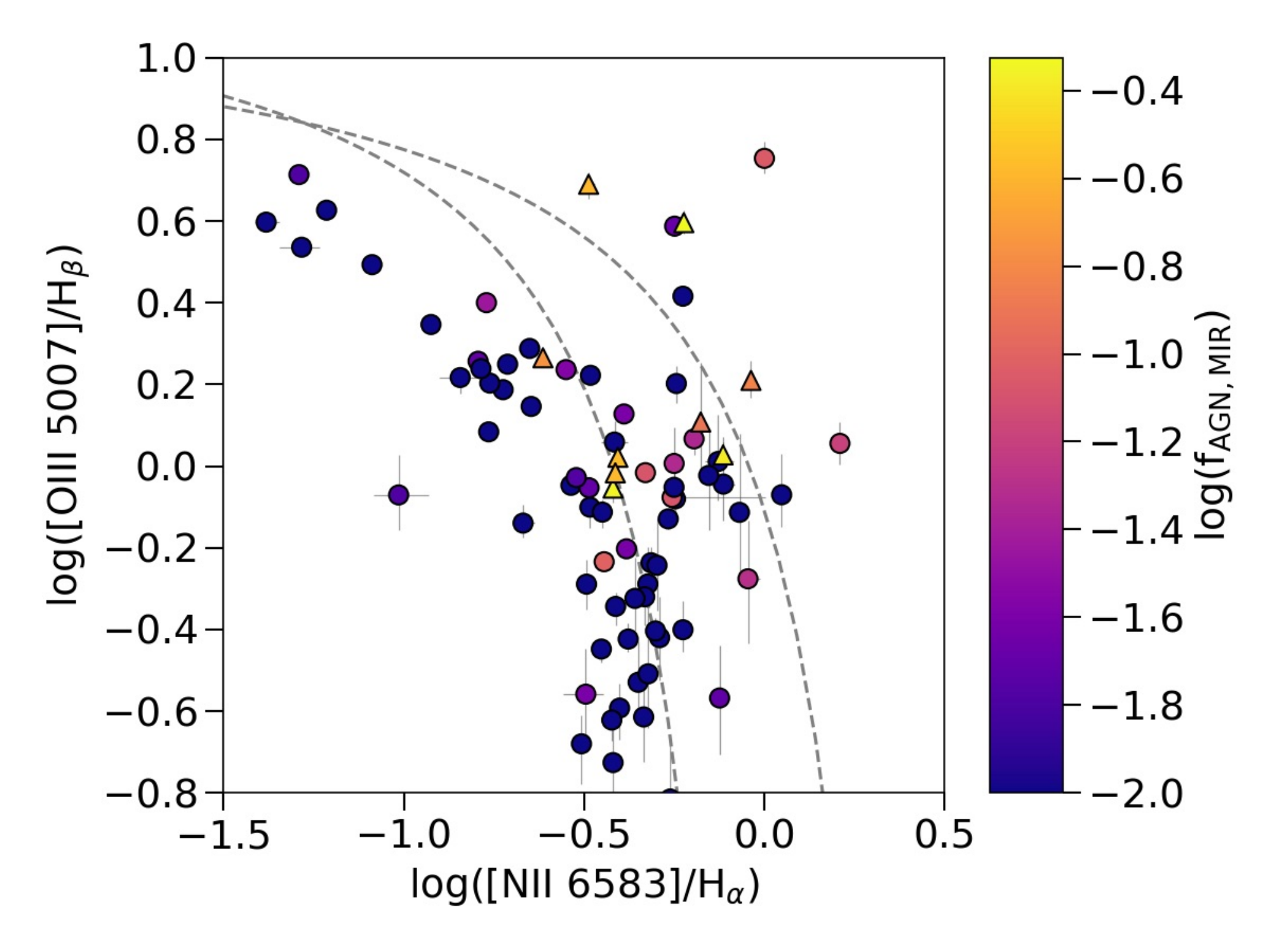}
\caption{Emission line diagnostics, with dashed lines from \citet{kewley06} denoting star-forming galaxies (lower left), composite galaxies (central region), and AGN/shocks (upper right). Triangles denote galaxies with \fmir{} $>0.1$ from the \mname{} fits to the photometry. SED-identified AGN are more likely to lie in the AGN regime than normal galaxies, though they are still found in all parts of the diagram.}
\label{fig:bpt}
\end{center}
\end{figure}

Figure \ref{fig:bpt} shows the \citet{brown14} sample on the BPT diagram. Only galaxies with observational line ratio errors of $<0.2$ dex are shown. There are 9 galaxies with \fmir{}$>$ 0.1: 3 lay in the AGN region, 4 are in the composite region, and 2 are in the star-forming region. The mild overlap between IR-selected AGN and AGN selected by their optical emission line ratios is consistent with results from the literature \citep{juneau13}.

\label{sec:bpt}

\subsection{Summary of AGN indicators}
\label{sec:agn_indicator}
\begin{figure}[th!]
\begin{center}
\includegraphics[width=0.95\linewidth]{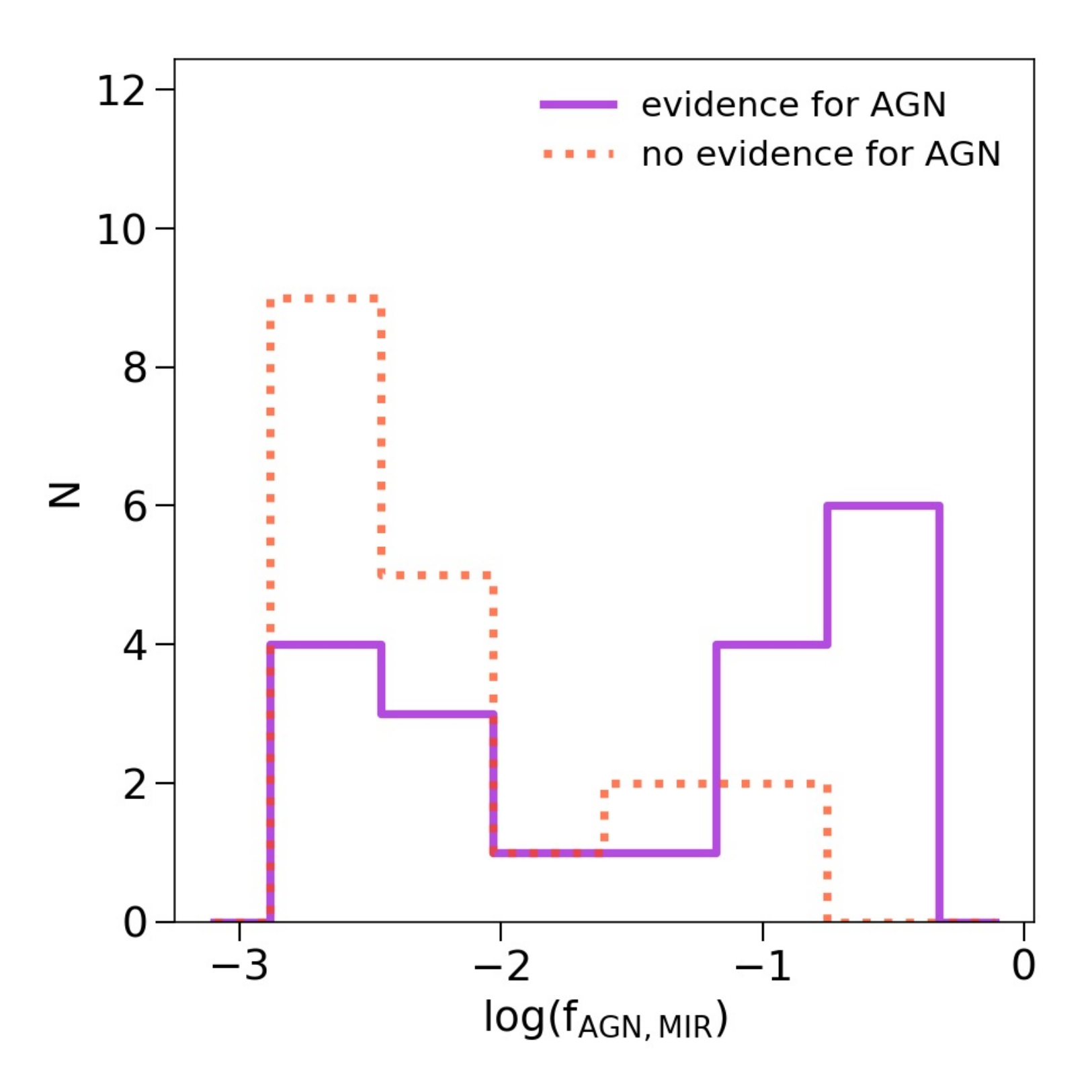}
\caption{Histogram of \fmir{} values, split into those with external evidence for AGN (purple) and those lacking external evidence for AGN (red). There is a clear correlation between \fmir{} and external evidence for AGN. Evidence for AGN constitutes any of the following: (a) an AGN emission line classification on the BPT diagram, (b) a \wisegradient{} of $<-0.15$ magnitudes/kpc, or (c) L$_{\mathrm{X}} > 10^{42}$ erg/s.}
\label{fig:agn_evidence}
\end{center}
\end{figure}

Figure \ref{fig:agn_evidence} shows the distribution of \fmir{} for galaxies with external evidence for AGN activity, and for galaxies with no external evidence for AGN activity. Evidence for AGN activity is classified as any of the following: (a) an AGN emission line classification on the BPT diagram, (b) a \wisegradient{} of $<-0.15$ magnitudes/kpc (chosen as the $\sim$10th percentile of the distribution of color gradients in the sample), or (c) L$_{\mathrm{X}} > 10^{42}$ erg/s. Nondetections are required to both (a) be covered in all three indicators, and (b) have no AGN evidence in all three indicators. The \wise{} integrated colors are not included as external AGN evidence here as they rely on the same broadband photometry from which \fmir{} is derived.

There is a clear separation in the histogram such that galaxies with high \fmir{} values are more likely to have AGN indicators at other wavelengths, and galaxies at low \fmir{} values are more likely to not have AGN indicators at other wavelengths. This strongly suggests that the hot dust emission parameterized by \fmir{} is indeed associated with buried AGN in this sample.

More broadly, however, the lack of a universal or complete indicator for AGN activity has proven challenging to building a complete census of AGN \citep{juneau13,trump15}. Fortunately, the \citet{brown14} sample consists of local galaxies which have collectively been well-studied in the extragalactic literature. In Table \ref{table:agnevidence}, we present a summary of the AGN indicators assembled in this work, and also highlight studies of these galaxies in the literature which shed light on the presence or absence of an AGN in these systems. This table includes all galaxies with \fmir{} $> 0.1$, and any galaxy with a positive AGN indicator.

This table highlights the careful, multiwavelength work currently necessary to identify a complete sample of AGN. Some of the non-detections in Figure \ref{fig:agn_evidence} with high \fmir{} values show other evidence for AGN activity. For example, NGC 0695 has \fmir{}= 0.097 and L$_{\mathrm{X}}$ = 10$^{40.8}$ erg/s from the X-ray catalogs consulted in this work, but more careful work including X-ray spectral fitting in \citet{brightman11} finds an AGN with L$_{\mathrm{X}}$ = 10$^{41.6}$ erg/s. IRAS 08572+3915 has the highest \fmir{} of the sample, \fmir{} = 0.87, and no clear evidence from the adopted AGN indicators, yet it is one of the brightest ULIRGs in the local universe and detailed SED modeling by \citet{efstathiou14} shows that up to 90\% of the bolometric luminosity of the system is emitted from the AGN alone. 

Conversely, there are multiple galaxies that meet the AGN evidence criteria but have low \fmir{} values. Four of these galaxies have significant \wise{} color gradients but otherwise no AGN evidence. \wise{} color gradients can be created by other astrophysical processes though, such as a radial gradient in dust heating. Others are BPT AGN but have no evidence for AGN activity in the mid-infrared, which is consistent with the non-overlapping selection techniques of AGN in the literature.

Overall, this literature search combined with the evidence presented in this section confirm that the large majority of galaxies with high \fmir{} values host AGN, and that the \citet{nenkova08b} formalism implemented in FSPS and \prospector{} is successful in detecting AGN in full-SED fits.

\begin{center}
\startlongtable
\begin{deluxetable*}{cccccp{50mm}}
\tablewidth{0pt}
\tablenum{1}
\tablecaption{Summary of AGN Evidence}
\label{table:agnevidence}
\tablehead{\colhead{Name} & \colhead{log(f$_{\mathrm{MIR,AGN}}$)} & \colhead{BPT} & \colhead{$\nabla$(W1-W2)} & \colhead{log(L$_X$)} & \colhead{Literature Notes} \\ \colhead{ }  & \colhead{ }  & \colhead{ }  & \colhead{mag/kpc}  & \colhead{erg/s}  & \colhead{ } } 
\startdata
IRAS 08572+3915 & ${-0.06}_{-0.02}^{+0.03}$ & --- & --- & 40.31 &  most luminous ULIRG in local universe, 90\% of L$_{\mathrm{bol}}$ from AGN \citep{efstathiou14}.\\
UGC 05101 & ${-0.18}_{-0.04}^{+0.04}$ & --- & --- & 40.45 &  Compton-thick ULIRG+AGN from X-ray spectra \citep{oda17}.\\
NGC 7674 & ${-0.33}_{-0.10}^{+0.10}$ & AGN & ${-0.1653}\pm{0.0001}$ & --- &  Seyfert 2 galaxy, continuum radio structure suggests AGN jets \citep{momjian03}.\\
CGCG 436-030 & ${-0.34}_{-0.10}^{+0.15}$ & star-forming & ${-0.1530}\pm{0.0008}$ & 40.68 &  Radio and PAH EW suggest AGN, composite / LINER BPT classification \citep{vardoulaki15}.\\
IC 5298 & ${-0.38}_{-0.11}^{+0.12}$ & composite & ${-0.2004}\pm{0.0004}$ & --- &  radio AGN, Seyfert 2 galaxy, and a LINER \citep{vardoulaki15}.\\
NGC 3690 & ${-0.56}_{-0.09}^{+0.11}$ & composite & ${-0.1679}\pm{0.0001}$ & 39.14 &  highly obscured AGN: N$_{\mathrm{H}} \sim 2.5 \times 10^{24}$ cm$^{-2}$) \citep{anastasopoulou16}.\\
NGC 7592 & ${-0.58}_{-0.16}^{+0.23}$ & composite & ${-0.2089}\pm{0.0009}$ & 40.90 &  Seyfert 2 galaxy \citep{maia03}.\\
UGC 08696 & ${-0.59}_{-0.14}^{+0.13}$ & AGN & --- & --- &  X-ray detection, obscured AGN N$_{\mathrm{H}} \sim 4 \times 10^{23}$ cm$^{-2}$ \citep{iwasawa11}.\\
IC 0691 & ${-0.74}_{-0.14}^{+0.11}$ & star-forming & --- & --- &  no further literature evidence for AGN.\\
IRAS 17208-0014 & ${-0.76}_{-0.19}^{+0.25}$ & --- & --- & 40.71 &  Compton thick AGN, strong MIR emission \citep{burillo15}.\\
NGC 1275 & ${-0.82}_{-0.13}^{+0.16}$ & AGN & ${-0.2435}\pm{0.0002}$ & 42.83 &  Seyfert 1 or 1.5, a BL Lac and an FR I radio source \citep{wilman05,panessa16}\\
IC 0883 & ${-0.90}_{-0.25}^{+0.50}$ & AGN & ${-0.0897}\pm{0.0004}$ & 39.87 &  starburst, radio emission confirms presence of AGN. \citep{romero12,romero17}\\
NGC 1144 & ${-0.93}_{-0.30}^{+0.92}$ & --- & ${-0.1027}\pm{0.0004}$ & --- &  Seyfert 2, with L$_{\mathrm{X}} \sim 10^{43.6}$ \citep{kawamuro16}.\\
NGC 0695 & ${-1.01}_{-0.27}^{+0.66}$ & star-forming & ${-0.0870}\pm{0.0005}$ & 40.80 &  low-obscuration AGN, careful analysis gives L$_{\mathrm{X}} \sim 10^{41.6}$ \citep{brightman11}\\
NGC 1068 & ${-1.04}_{-0.52}^{+1.93}$ & AGN & ${-0.3610}\pm{0.0001}$ & 41.21 &  classic Compton-thick AGN identified in Xrays \citep{lopezgonzaga16}\\
NGC 4194 & ${-1.08}_{-0.31}^{+0.35}$ & composite & ${-0.1398}\pm{0.0009}$ & 40.13 &  AGN according to X-ray and MIR diagnostics \citep{lehmer10}\\
UGC 09618 & ${-1.08}_{-0.30}^{+0.93}$ & AGN & --- & 40.86 &  radio AGN \citep{vardoulaki15}\\
NGC 6240 & ${-1.18}_{-0.45}^{+1.72}$ & AGN & ${-0.1053}\pm{0.0002}$ & --- &  bright X-ray AGN (L$_{\mathrm{X}} \sim 10^{43.5}$) \citep{brightman11}\\
NGC 2623 & ${-1.29}_{-0.47}^{+1.40}$ & composite & ${-0.1305}\pm{0.0011}$ & 40.84 &  classified as AGN by radio and MIR PAH EW \citep{vardoulaki15}\\
NGC 6090 & ${-1.60}_{-0.65}^{+1.58}$ & star-forming & ${-0.0460}\pm{0.0101}$ & 40.61 &  radio AGN, optical Seyfert, LIRG \citep{vardoulaki15}\\
NGC 5256 & ${-1.68}_{-0.76}^{+1.64}$ & AGN & ${-0.0953}\pm{0.0005}$ & 41.62 &  radio AGN, no AGN signature in MIR continuum or PAH EW \citep{imanishi10,vardoulaki15}\\
Arp 256 S & ${-1.69}_{-0.62}^{+1.62}$ & star-forming & ${-0.1308}\pm{0.0010}$ & 40.15 &  starburst with high PAH EW (no AGN) \citep{stierwalt13}\\
NGC 4579 & ${-2.09}_{-0.99}^{+1.16}$ & AGN & ${-0.0523}\pm{0.0005}$ & 41.15 &  low luminosity (L$_{\mathrm{X}}=10^{41}$ erg/s), low obscuration (log(N$_{\mathrm{H}}$)=20.6) AGN \citep{asmus15}\\
III Zw 035 & ${-2.19}_{-0.85}^{+1.39}$ & composite & ${-0.1528}\pm{0.0016}$ & 40.06 &  starburst in radio \citep{vardoulaki15}\\
NGC 4450 & ${-2.25}_{-0.94}^{+1.03}$ & composite & ${-0.0019}\pm{0.0006}$ & 40.39 &  potential low-luminosity AGN (L$_{\mathrm{X}}=10^{40.5}$) \citep{liu11}\\
NGC 7673 & ${-2.30}_{-1.04}^{+1.18}$ & star-forming & ${-0.0038}\pm{0.0015}$ & 40.09 &  no further literature evidence for AGN\\
NGC 4138 & ${-2.32}_{-0.99}^{+1.09}$ & composite & ${-0.0390}\pm{0.0013}$ & 40.25 &  potential low-luminosity AGN (L$_{\mathrm{X}}=10^{40.7}$) \citep{chen17}\\
UGCA 208 & ${-2.37}_{-1.01}^{+1.11}$ & AGN & --- & --- &  low-mass interacting galaxy, no other AGN indicators. BPT status from shocks? \citep{smith10}\\
NGC 4670 & ${-2.41}_{-0.97}^{+1.07}$ & star-forming & ${-0.0909}\pm{0.0054}$ & 39.76 &  no further literature evidence for AGN\\
NGC 3079 & ${-2.45}_{-0.99}^{+1.27}$ & composite & ${-0.0694}\pm{0.0003}$ & 39.76 &  luminous in hard X-rays, potential low-luminosity Compton-thick AGN \citep{ricci15}\\
Arp 256 N & ${-2.46}_{-1.07}^{+1.27}$ & star-forming & ${-0.0202}\pm{0.0042}$ & 40.16 &  no further literature evidence for AGN\\
NGC 3310 & ${-2.47}_{-0.98}^{+1.23}$ & star-forming & ${-0.0103}\pm{0.0007}$ & 39.96 &  no further literature evidence for AGN\\
NGC 4569 & ${-2.52}_{-1.04}^{+0.98}$ & AGN & --- & 39.85 &  optical AGN, but with low L$_{\mathrm{X}}$ and L$_{\mathrm{MIR}}$ \citep{alonsoherrero16}\\
NGC 0520 & ${-2.56}_{-1.12}^{+1.27}$ & --- & ${-0.1910}\pm{0.0005}$ & 40.19 &  no further literature evidence for AGN\\
NGC 4676 A & ${-2.64}_{-1.14}^{+1.10}$ & composite & ${-0.0995}\pm{0.0008}$ & 39.90 &  no further literature evidence for AGN\\
NGC 2798 & ${-2.65}_{-1.02}^{+1.06}$ & composite & ${-0.1070}\pm{0.0018}$ & 39.67 &  no further literature evidence for AGN\\
NGC 1614 & ${-2.67}_{-0.99}^{+1.19}$ & composite & ${-0.1712}\pm{0.0004}$ & 39.85 &  no further literature evidence for AGN\\
NGC 5033 & ${-2.69}_{-1.06}^{+1.04}$ & composite & ${-0.0023}\pm{0.0004}$ & 41.02 &  weak X-ray AGN (L$_{\mathrm{X}}=10^{41}$) \citep{asmus15}\\
NGC 5953 & ${-2.74}_{-1.18}^{+1.17}$ & composite & ${-0.0951}\pm{0.0009}$ & 39.97 &  Seyfert Type 2 \citep{tsai15}\\
NGC 7714 & ${-2.76}_{-0.95}^{+1.02}$ & composite & ${-0.1177}\pm{0.0013}$ & 40.81 &  no further literature evidence for AGN\\
IC 4553 & ${-2.78}_{-0.98}^{+1.07}$ & --- & ${-0.2233}\pm{0.0004}$ & --- &  also known as Arp 220. AGN presence unknown (see e.g. \citealt{paggi17})\\
NGC 4254 & ${-2.88}_{-0.99}^{+0.89}$ & star-forming & ${0.0285}\pm{0.0004}$ & 39.03 &  no further literature evidence for AGN\\
Mrk 33 & ${-2.88}_{-0.96}^{+0.90}$ & star-forming & ${-0.1119}\pm{0.0070}$ & 39.08 &  no further literature evidence for AGN\\
\enddata
\tablecomments{Galaxies included in this table fulfill one or more of the following criteria: (1) meet the independent AGN criteria described in Section \ref{sec:agn_indicator}, (2) have \fmir{} $> 0.1$ from the \mname{} fits, or (3) have all three AGN criteria measured, and do not meet any of them. All L$_{\mathrm{X}}$ values in the table above are in units of erg/s.}
\end{deluxetable*}
\end{center}

\section{Discussion}
\label{sec:discussion}
\subsection{The utility of MIR AGN templates in UV-IR SED fits}
\label{sec:bigdisc}
Most galaxy SED fitting routines do not incorporate emission from AGN in their physical models. Thus, a key question is to what extent this machinery is necessary for fitting the SEDs of a general population of galaxies. Here, we discuss the factors relevant to answering this question.

This work has made it clear that when an IR AGN is present in a galaxy and MIR data are present, it is necessary to include an AGN emission model in order to recover accurate SFRs, SFHs, dust attenuations, and metallicities. Simply masking mid-infrared data is an inadequate solution: without expensive far-infrared or sub-millimeter photometry, mid-infrared emission is the only way to measure the dust emission of galaxies. Constraints on the IR luminosity are critical to measure accurate star formation rates: \citet{fang17} show that UV+optical SED-based SFR estimates for the general galaxy population show 0.2 dex systematics which vary as a function of mass and redshift, while \citep{wuyts11a} show that optical to near-infrared SED-based SFRs can be biased low by 0.5 dex or more for highly star-forming galaxies. Furthermore, dust optical depths estimated from SED fits can be systematically underestimated by up to 0.5 for typical star-forming galaxies without IR data, though this is strongly dependent on the adopted priors \citep{leja17}.

Another important question is the prevalence of AGN in the general galaxy population. In the \citet{brown14} sample investigated in this paper, approximately 10\% of galaxies have high \fmir{} values; however, this result cannot be generalized to the entire galaxy population as the \citet{brown14} sample is not volume-complete. \citet{kartaltepe10} find that 15\% of star-forming galaxies host IR AGN, according to classic IR color-color selection criteria \citep{stern12}. This color-color criteria will identify AGN which dominate over the MIR emission of their host galaxy: however, it will not identify low-luminosity IR AGN, which are likely to have a more subtle effect on SED fits. Using a more complete set of diagnostics, including X-rays, IR, radio, and the mass-excitation diagram, \citet{juneau13} find that 37\% of star-forming galaxies at $0.3 < z < 1.0$ host AGN. \citet{kauffmann03b} use BPT diagnostics in the SDSS and find that 35\% of SDSS sources with all 4 measured lines are classified as composite or AGN, while 13\% are classified as AGN alone. Clearly, the preponderance of evidence from the literature suggests that a significant minority of galaxies host AGN. 

The importance of an AGN dust emission model is likely to increase for galaxies at higher redshifts, as AGN accretion rates are thought to peak at $z \sim 2$ \citep{richards06,hopkins07}. At higher redshifts, \citet{kirkpatrick15} find that up to 40\% of 24$\mu$m-selected galaxies host an AGN, even for faint selection thresholds. \citet{marsan15,marsan17} find $>80\%$ of very massive galaxies at $3 < z < 4$ have MIR emission which is significantly contaminated or dominated by AGN. \citet{messias13} find that the hot dust luminosity ($\sim$3-4$\mu$m) increases much faster with increasing redshift than cold dust luminosity, and that some $\sim30\%$ of hot dust luminosity is attributed to AGN activity.

Another important factor is the MIR photometric coverage necessary to distinguish between AGN dust emission and dust heated by star formation. The \citet{nenkova08b} AGN templates adopted in this work have MIR shapes which are distinct from the \citet{draine07} templates governing the rest of the IR SED (see Figure \ref{fig:wise_colors}): as a result, mock tests show that \fmir{} is uniquely determined even with 3-4 bands of IR photometry. In the limit of 1-2 bands of IR photometry, however, the priors in the galaxy SED model will play a larger role in determining the source of the dust emission. Even in this limiting case, it is important to include an AGN template so that the model error bars accurately reflect the lack of knowledge about what is powering the dust emission of the galaxy. The priors must be carefully tuned such that they do not blindly attribute all dust emission to star formation or AGN, but instead reflect the best estimates of the distribution of galaxy and AGN luminosities in the target galaxy population.

In summary, it is recommended that the template for MIR AGN emission be adopted in the majority of galaxy SED fits that include MIR photometry, unless there is strong reason to believe that the target population does not host AGN. We further recommend careful tuning of the Bayesian priors in the SED model to accurately reflect the prevalence and strength of AGN in the target population, especially when the available IR information is limited.

\subsection{Other sources of hot dust emission}
\label{sec:hotdust}
The improvement in the MIR photometric residuals and in the predicted MIR spectra implies that hot dust is a necessary component of SED fitting. The preponderance of evidence presented in Section \ref{sec:agn} suggests that this hot dust is indeed associated with an AGN. However, we cannot rule out the possibility that \fmir{} is also sensitive to non-AGN sources of hot dust emission.

One possibility is that the AGN component in \mname{} is actually being fit to dust heated to high temperatures by extreme star formation. Given that many of the \mname{} AGN are detected in ULIRGS, which are likely powered by intense, dust-obscured starbursts, this is not an unreasonable hypothesis. The \citet{draine07} dust emission model used in \mname{} does not specifically contain a separate hot dust component, but instead contains emission at range of temperatures based on a physical model for dust heated by starlight. This model has been shown to be a reasonable representation of resolved star-forming regions in nearby galaxies \citep{aniano12}, where even pure photodissociation regions have \citet{draine07} parameters which are not near the extremes of the \citet{draine07} parameter space. It is thus unlikely that strong radiation field in these galaxies can be produced by star formation alone: a significant fraction of the volume in these galaxies would have to have more extreme conditions than an active HII region. Furthermore, this extreme star formation would have to be relatively common, occurring in 10\% of the \citet{brown14} sample. Notably, however, SED models may disagree on this point; while the most extreme set of \citet{draine07} parameters corresponds approximately to 45 K, other SED models include hot dust emission associated with star formation at 130 K and 250 K \citep{dacunha08}.

It is also possible that the \citet{nenkova08b} AGN template picks up MIR emission which is powered by neither star formation nor AGN activity, but instead by some other physical mechanism. It has long been known that star-forming galaxies show excess emission at $3-5\mu$m both in the local universe \citep{imanishi00,mentuch09} and at higher redshifts \citep{magnelli08,lange16}. Multiple ideas have been put forth for the origin of this emission, including an unusually strong 3.3$\mu$m PAH feature \citep{magnelli08} and circumstellar disk emission from intermediate-aged stars \citep{mentuch09,meidt12}. However, neither of these hypotheses can adequately explain the $3-5\mu$m emission observed in these galaxies: {\it Akari} spectra suggest that \mname{} is predicting the strength of the 3.3$\mu$m PAH feature with reasonable accuracy (Figure \ref{fig:resid}), and the \mname{} model includes hot dust emission from intermediate-aged stars. Furthermore, we have demonstrated that the majority of hot dust emission is coming from a central source (Section \ref{sec:wisegradients}). While we cannot rule out a small contribution to \fmir{} coming from diffuse emission which is unrelated to AGN or star formation activity, it is unlikely to be a dominant component.

\subsection{Comparing to other IR AGN selection methods}
\label{sec:other_techniques}
While the AGN component of \mname{} is primarily built to improve the accuracy of galaxy properties in galaxies that host AGN, it can also be used to identify AGN in galaxies. Here we briefly sketch how \mname{} fits into the constellation of IR AGN identification methods based on broadband photometry.

Many studies in the literature use integrated MIR colors to identify AGN \citep{stern05,goulding09,jarrett11,stern12,assef13}. The advantage of this approach is the simplicity: it requires only observed MIR photometry and does not depend on the details of any given AGN emission model. The disadvantage is also the simplicity. Intrinsically, the observed photometry measures a mixture of galaxy and AGN emission, and simple color criteria return only a binary for presence or lack of AGN. This forces a choice between ``completeness" and ``contamination" when selecting AGN with broadband colors, where potential contaminants include star-forming galaxies with strong PAH emission \citep{stern05,assef10,hainline16}, high-redshift massive galaxies \citep{donley07,donley12}, and non-galactic sources such as brown dwarfs and young stellar objects \citep{stern07,koenig12}.

The SED-fitting method presented in this work contains model components to deal with the potential galactic contaminants, and furthermore, returns a probability distribution function for AGN emission luminosities rather than a binary decision. This more sophisticated modeling approach can be used to identify AGN to lower luminosities than broadband color selection. For example, Figure \ref{fig:wise_colors} shows the \citet{brown14} sample on a \wise{} color-color plot: only 6 galaxies would be identified as hosting IR AGN with the selection criteria of \citet{stern12}, whereas \mname{} identifies 13 galaxies with a strong AGN component.

Other SED-fitting codes that simultaneously model galaxy and AGN emission have been developed. Many are built with the primary purpose of identifying AGN in different wavelength regimes: at infrared wavelengths \citep{sajina06,han12,hernan15,suh17}, simultaneously in the UV, optical, and infrared \citep{rivera16}, or for UV-IR and radio wavelengths \citep{ciesla15b}. These codes use relatively simple models for galaxy emission, and often do not model complex variations in galaxy physical properties (e.g., varying PAH contribution, variations in the dust attenuation curve, variable stellar metallicity) which set the IR energy budget of galaxies \citep{leja17}. Codes that have both more complex variations in galaxy properties and AGN components include MAGPHYS \citep{dacunha08,berta13} and BayeSED \citep{han14}. AGN identification via panchromatic SED fitting shows great promise for disentangling the complexities of mixed star formation and AGN indicators \citep{chang17}, though this technique is not yet widespread in the literature.

When MIR spectra are available, it is also possible to use empirical libraries and spectral decomposition to diagnose the presence of a MIR AGN \citep{mullaney11,kirkpatrick12,sajina12,kirkpatrick15}. This empirical approach allows the user to separate MIR spectrum into an AGN component and a galaxy component, avoiding the binary classification noted above, and is also not dependent on the accuracy of MIR emission models. However, this approach is reliant on the accuracy of the library of observations on which it is built, and also subject to the selection criteria of these observations.

The advent of the James Webb Space Telescope (JWST) is likely to greatly increase the community demand for MIR AGN models in SED fitting, as it will provide high-resolution MIR galaxy photometry and spectra. SED models with MIR AGN components such as the one presented in this paper will be necessary to properly model and interpret these data. The development of SED models with MIR AGN components may still contain model-dependent systematics: \citet{roebuck16} fit radiative transfer simulations of dusty AGN emission and find that the systematic uncertainties in separating galaxy and AGN emission in the infrared are substantial, with 1$\sigma$ f(AGN)$_{\mathrm{IR}} \sim 0.4$. JWST data will provide a key testing ground for the uncertainties in these models.

\section{Conclusions}
\label{sec:conclusion}
In this work, we have incorporated AGN torus emission models from \citet{nenkova08b} into the \mname{} galaxy SED model. We have fit photometry from the \citet{brown14} spectrophotometric catalog with the \mname{} model twice, once with AGN and once without AGN emission. We show that galaxies which host AGN in the \mname{} model tend to live in star-forming galaxies on or above the star-forming sequence. We use these models fit to the photometry to predict optical and mid-infrared spectra from \citet{brown14}, which are {\it not} fit, and demonstrate that models fit with AGN templates are in much better agreement with the observed spectra. Additionally, we assess the quality of the derived star formation histories and stellar metallicities for galaxies with significant AGN components. In these galaxies (i.e., where \fmir{} $> 10\%$, which encompasses 10\% of the sample), we find the following improvements:

\begin{enumerate}
\item Residuals in the MIR photometry and in the MIR spectra show substantial improvement when the AGN templates are turned on, where the average improvement is between 20\%-60\% of the flux. For specific galaxies, the residuals can improve by up to a factor of 10.
\item The \halpha{} and \hbeta{} fluxes predicted from the physical model improve greatly with an AGN component, with scatter in both decreasing by $\sim$0.25 dex, suggesting a corresponding improvement in the model SFR and dust properties.
\item The prediction of the Balmer decrements from the dust reddening in the \mname{} physical model greatly improves. The offset in reddening decreases from 0.1 dex to $\sim$0.0 dex, and the scatter decreases from 0.15 dex to 0.1 dex.
\item Derived stellar metallicities are much more consistent with the stellar mass--stellar metallicity relationship from \citet{gallazzi05}, changing from a systematic offset of 2.3$\sigma$ to a systematic offset of 0.9$\sigma$.
\item Unphysical behavior in the derived star formation histories are removed. These biases made the derived stellar ages too low by up to a factor of ten. These were caused by circumstellar AGB dust mimicking AGN signatures.
\end{enumerate}

We also analyze whether the 3-8$\mu$m in galaxies with high \fmir{} is truly from dust heated to high temperatures by AGN, as opposed to an alternative source of dust heating. We demonstrate that \fmir{} is similar to classic \wise{} color-color cuts for AGN, and that the \mname{} fit identifies more AGN than color-color techniques. This is due to both the use of probabilistic approach and a full SED model for the dust emission from the host galaxy. We further assemble multiple lines of evidence to assess the origin of the hot dust emission picked up by the \citet{nenkova08b} AGN templates, including \wise{} MIR color gradients, {\it Chandra} X-ray luminosities, BPT diagnostics, and evidence from the literature ranging from radio emission to PAH equivalent widths. Given the internal disagreement amongst the indicators, they correlate well with \fmir{}, suggesting that our technique is primarily picking up hot dust emission from AGN. We note that we cannot rule out the contribution of diffuse hot dust emission at a low level.

Future work will focus on the ongoing question of to what extent the hot dust emission is linked to star formation, and the contribution to the SED from IR AGN at higher redshifts. These questions may be critical to determining accurate star formation rates from full galaxy SED fits.

\acknowledgements

We thank Josh Speagle, Philip Rosenfield, and Phillip Cargile for valuable discussion and comments. J.L. is supported by an NSF Astronomy and Astrophysics Postdoctoral Fellowship under award AST-1701487. The computations in this paper were run on the Odyssey cluster supported by the FAS Division of Science, Research Computing Group at Harvard University.

\bibliography{/Users/joel/my_papers/tex_files/jrlbib}
\end{document}